# Remote Moiré Modulation of Decoupled Dirac Subsystems in Twisted Trilayer Graphene

Dohun Kim[1,†], Junsik Choe[1,†], Minseo Cho[1], Takashi Taniguchi[2], Kenji Watanabe[3], Gil Young Cho[4,5], and Youngwook Kim[1,*]

[1]*Department of Physics and Chemistry, Daegu Gyeongbuk Institute of Science and Technology (DGIST), Daegu 42988, Republic of Korea*

[2]*Research Center for Materials Nanoarchitectonics, National Institute for Materials Science, 1-1 Namiki, Tsukuba 305-0044, Japan*

[3]*Research Center for Electronic and Optical Materials, National Institute for Materials Science, 1-1 Namiki, Tsukuba 305-0044, Japan*

[4]*Department of Physics, Korea Advanced Institute of Science and Technology, Daejeon 34141, Republic of Korea*

[5]*Center for Artificial Low Dimensional Electronic Systems, Institute for Basic Science, Pohang 37673, Korea*

[†]These authors contributed equally

[*]E-mail: y.kim@dgist.ac.kr

**ABSTRACT: Moiré superlattices are generally assumed to act only at the interface where lattice mismatch or twist occurs. Here, we study charge transport in large-angle helical twisted trilayer graphene, where interlayer tunneling is strongly reduced. When only the top monolayer graphene is aligned with hBN, the electronic response reorganizes into a moiré-modulated monolayer and a remaining twisted bilayer graphene subsystem. Despite the absence of any explicit structural moiré in the twisted bilayer, we observe satellite-like features in its electronic response that run parallel to the primary spectrum and are locked to the density scale of the hBN/graphene moiré. These observations are consistent with a weak moiré-periodic electrostatic perturbation generated at the aligned interface and sensed by a spatially separated Dirac subsystem, even in the absence of coherent interlayer coupling.**

## INTRODUCTION

Moiré superlattices in van der Waals heterostructures have enabled a range of emergent quantum phases, including flat-band superconductivity in twisted graphene systems and correlated states in moiré materials.[1–18] These phenomena arise because long-wavelength periodic potentials reorganize the electronic spectrum and interactions in two-dimensional systems. Since moiré potentials originate from local atomic registry at an interface, they are generally regarded as interfacial in nature, acting primarily on the layers directly involved in lattice mismatch or twist. Within this conventional picture, layers that are electronically decoupled from the moiré interface should remain largely unaffected, particularly when interlayer hybridization is suppressed.

Recent observations indicate that moiré-related effects need not be strictly confined to the immediate interface where lattice alignment occurs. For instance, correlated charge configurations in transition metal dichalcogenide moiré heterobilayers have been shown to generate periodic electrostatic potentials that influence nearby monolayers.[19–21] In rhombohedral graphene aligned to hexagonal boron nitride (hBN), scanning-probe measurements have revealed moiré-induced features extending deeper into the multilayer stack,[22] which are essential for the emergence of correlated phases such as the fractional and integer quantum anomalous Hall effect.[23–28] In these systems, however, the propagation of moiré signatures is generally accompanied by substantial interlayer tunneling or structural reconstruction. Disentangling purely electrostatic modulation from hybridization-driven effects therefore remains challenging, and the spatial reach of a moiré potential in electronically decoupled Dirac multilayers is not yet fully established.

Large-angle helical twisted trilayer graphene (TTG) provides a platform in which this issue can be addressed in a controlled regime. At twist angles exceeding approximately 2°, the momentum mismatch between adjacent layers strongly suppresses interlayer hybridization,[29–37] and intrinsic miniband features are pushed away from the low-energy window relevant for transport. The system can thus be viewed as a stack of electronically decoupled Dirac layers that remain capacitively coupled, as described by a multilayer electrostatic model. When only the top monolayer graphene is aligned to hBN, an hBN/graphene moiré superlattice is formed at the top interface, while the middle and bottom layers constitute a twisted bilayer graphene (TBG) subsystem without structural moiré (Fig. 2c). Within a conventional local-screening picture, the electronic spectrum of this remote TBG block would be expected to be largely insensitive to a periodic potential confined to the top interface. This geometry therefore enables an examination of whether density scales associated with a moiré-modulated monolayer can manifest in a spatially separated Dirac subsystem in the absence of strong coherent interlayer tunneling.

Here we report transport measurements on large-angle helical TTG in which only the top monolayer is aligned to hBN. The aligned top layer forms an hBN/graphene moiré, while the remaining two layers form an underlying twisted bilayer subsystem that lacks a structural moiré. In this setting, satellite-like features appear in the TBG response at carrier densities set by the hBN/graphene moiré and remain visible down to $B = 0$. By systematically investigating three devices with hBN/top graphene twist angles of approximately 0°, 1°, and 2°, we find that the carrier density at which the satellite features of the remaining TBG subsystem emerge depends on the twist angle. These features are consistent with a scenario in which the hBN/top-graphene moiré induces a periodic density modulation in the top monolayer, which in turn generates a moiré-periodic electrostatic potential that propagates through the trilayer stack and is sensed by the lower subsystem. Notably, this occurs in a regime where coherent interlayer tunneling is strongly suppressed, suggesting that in other moiré materials, such as rhombohedral multilayer graphene, stronger interlayer tunneling and hybridization may further enhance the spatial reach of moiré-induced modulation.[24–26]

## RESULTS AND DISCUSSION

### Configuration dependence in twisted trilayer graphene

Figure 1 compares the transport response of two helical TTG devices measured using dual graphite gates. The longitudinal conductance $\sigma_{xx}$ is mapped at $B = 3$ T as a function of the total carrier density, $n_{tot} = (C_t V_t + C_b V_b)/e$, and the displacement field, $D/\varepsilon_0 = (C_t V_t - C_b V_b)/(2\varepsilon_0)$, where $C_t$ and $C_b$ are the gate capacitances determined by the thickness of the hBN dielectrics and $V_t$ and $V_b$ are the applied top and bottom gate voltages. Device H1 serves as a reference without hBN/graphene alignment, whereas device M1 is structurally similar except that only the top monolayer graphene is aligned with the adjacent hBN, forming an hBN/graphene moiré superlattice. In both devices, the relative twist angles between adjacent layers are approximately 5°. In the absence of hBN/graphene alignment (device H1), the conductance map reveals multiple sets of Landau-level features, including diagonal trajectories with opposite slopes and nearly vertical structures. These branches can be assigned to the top, middle, and bottom graphene layers, showing that the trilayer behaves as three effectively decoupled Dirac subsystems.

The Landau levels associated with the outer layers show a pronounced dependence on the displacement field, whereas those of the middle layer respond more weakly due to screening by the outer layers. The diagonal trajectories with opposite slopes directly reflect layer polarization: the Landau levels of the top graphene layer are primarily coupled to the top gate and disperse with displacement field in one direction (green dotted line in Fig. 1a), while those of the bottom layer are mainly controlled by the bottom gate and disperse in the opposite direction (red dotted line in Fig. 1a).[38,39] This separation of gate response provides an experimental reference in which TTG behaves as three largely independent Dirac sheets in the absence of a moiré perturbation. A similar response is also observed in a large-angle alternating TTG (device A) and an additional helical TTG (device H2) (Supplementary Fig. S2 and Supplementary Note 1), consistent with the decoupling expected outside the small-angle regime where strong interlayer hybridization dominates.

By contrast, when only the top monolayer graphene is aligned with the top hBN (device M1), the conductance map exhibits a markedly simplified structure. In this device, the relative twist angles between adjacent graphene layers are both approximately 5°, placing the trilayer well within the large-angle regime. The top and bottom graphene layers therefore differ in orientation by about 10°, and the bottom graphene layer is intentionally misaligned with respect to the lower hBN. As a result, a structural moiré superlattice is formed only at the top interface, while the bottom interface remains moiré-free.

Compared to device H1, where three features are resolved, device M1 exhibits only two: a single diagonal Landau-level branch coexisting with one nearly vertical feature. The disappearance of one diagonal branch, while the vertical feature remains, signals a redistribution of the electronic response following hBN/graphene alignment. Because the top alignment is the only intentional structural modification relative to device H1, this contrast shows that hBN/graphene alignment alters the experimentally resolved subsystems in the large-angle regime, without invoking flat-band formation. The microscopic origin and layer assignment of the remaining features are examined through a subsystem-resolved transport analysis in the following sections.

**Two subsystems in twisted trilayer graphene with a moiré potential**

Figure 2 presents contour maps of $\sigma_{xx}$ in device M1 measured at $B$ = 2 T and 1 T as functions of total density $n_{tot}$ and displacement field $D/\varepsilon_0$. Two types of quantum Hall features are resolved: diagonal Landau-level branches with non-uniform spacing characteristic of monolayer graphene, and a nearly vertical feature that persists over a wide range of displacement fields. The coexistence of these diagonal and vertical responses indicates that the trilayer is no longer described by three equivalent Dirac layers, but instead separates into two electronic subsystems. In the following, we assign the diagonal branches to a monolayer graphene subsystem and the vertical feature to a TBG subsystem. Here and throughout, TBG subsystem refers to the middle–bottom twisted bilayer as a structural and electrostatic unit; at $\theta \approx 5°$ its single-particle spectrum consists of two nearly decoupled Dirac cones, and the term does not imply hybridized moiré minibands.

The vertical feature exhibits its own filling-factor sequence and, at $B$ = 2 T, progresses in steps of $\Delta\nu_{tot} = 4$. For a TBG subsystem with preserved layer symmetry, an eightfold sequence would be expected from spin, valley, and layer degeneracies. The observed fourfold sequence therefore indicates that the layer degeneracy of the TBG subsystem is lifted. In device M1, the middle–bottom bilayer does not reside at zero displacement field. Owing to the dual-gate configuration and electrostatic screening by the aligned top monolayer graphene, a finite displacement field is generally present across the TBG block under typical gating conditions. Such an interlayer potential difference lifts the layer degeneracy at the onset of Landau quantization and reduces the eightfold sequence to fourfold, consistent with the observed vertical feature.

Nevertheless, the middle and bottom graphene layers do not appear as two separately resolved monolayer branches, but instead produce a single TBG-like response. This can be understood by noting that the spatially uniform interlayer potential difference responsible for lifting the layer degeneracy is distinct from the moiré-periodic perturbation transferred from the top graphene. The former corresponds to the zero-wavevector ($q$=0) component of the electrostatic potential, whereas the latter varies spatially with the finite moiré wavevector ($G_m$).

The hBN/graphene moiré potential induces a periodic carrier-density modulation in the aligned top graphene, which in turn generates a Hartree potential with the same periodicity. This periodic potential penetrates into both the middle and bottom graphene layers. Our electrostatic calculation shows that the amplitudes experienced by the two lower layers are comparable, so that the perturbation is dominated by the symmetric combination of the middle- and bottom-layer potentials, with only a weak antisymmetric component. Because this symmetric mode couples primarily to the total lower-layer density, $n_{mid} + n_{bottom}$, the two layers respond collectively to the moiré-periodic perturbation. This predominantly symmetric coupling is consistent with the nearly vertical TBG-like feature, particularly evident in Fig. 2a, and provides indirect support for the electrostatic transfer of the moiré modulation from the aligned top graphene to the lower bilayer subsystem. Further details are provided in Supplementary Note 3.

In addition, the slopes of the diagonal branches change abruptly as they cross the vertical feature (cyan dotted line), signaling a redistribution of charge between the two subsystems. This slope change reflects the fact that the partitioning of the total carrier density between the monolayer graphene and the TBG block depends on the filling of the TBG subsystem. When the TBG subsystem is tuned to its charge neutrality point, variations in the total carrier density are accommodated predominantly by the monolayer graphene subsystem.

Following approaches established in monolayer-on-TBG and sensor-layer geometries, this slope change enables extraction of the interlayer capacitance, $C_{int}$, between the monolayer graphene and the TBG subsystem: at the TBG charge neutrality point, matching the monolayer chemical potential to its Landau-level index determines $C_{int}$. For device M1, two layer assignments are, in principle, possible: (i) a top monolayer graphene coupled to a middle–bottom TBG block, or (ii) a bottom monolayer graphene coupled to a top–middle TBG block. Under configuration (ii), the extracted $C_{int}$ takes a negative value within the experimental parameter range, which is inconsistent with a physically meaningful interlayer capacitance. In contrast, configuration (i) yields $C_{int}$ = 6.1 μF cm$^{-2}$, comparable to values reported for monolayer graphene/magic-angle TBG sensor-layer systems[35] and consistent with the known device geometry in which the top monolayer graphene is aligned with the adjacent hBN. Details of the extraction procedure are provided in Supplementary Note 2.

The diagonal branches further exhibit signatures of symmetry breaking within the monolayer subsystem. In Fig. 2, the $N = 0$ Landau level shows a clear splitting at $B = 2$ T and 1 T, indicating lifted degeneracy. Tracking this splitting to lower magnetic fields reveals that it persists continuously down to zero field (Supplementary Fig. S7). The persistence of the $N = 0$ splitting at $B = 0$ is consistent with valley symmetry breaking in the aligned monolayer, as expected for an hBN/graphene moiré that breaks the sublattice symmetry of graphene.

These observations can be understood in the context of the hBN/graphene moiré reconstructing the Dirac spectrum of the aligned top monolayer and redistributing spectral weight within the moiré Brillouin zone.[40–44] Such reconstruction suppresses momentum-conserving tunneling processes across the top–middle interface in the low-energy regime. By contrast, the middle and bottom graphene layers preserve a Dirac-like spectrum within the energy window probed here and continue to form a bilayer subsystem. The trilayer can therefore be described, at low energies, as a moiré-modulated monolayer coexisting with a remaining twisted-bilayer block.

**Moiré-induced density modulation and satellite features in the lower subsystem**

Having established that hBN alignment isolates a moiré-modulated monolayer from the remaining TBG block, we next characterize the density scale associated with the hBN/graphene moiré in the aligned layer. In hBN/graphene heterostructures, the moiré superlattice generates a long-wavelength periodic potential that reconstructs the Dirac spectrum of the monolayer, producing secondary Dirac points at well-defined carrier densities that persist independently of magnetic field.

We first determine this characteristic density scale in device M1. The derivative maps $d\sigma_{xx}/dn_{tot}$ in Fig. 3a and b reveal a secondary Dirac point associated with the hBN/graphene moiré superlattice that persists down to zero magnetic field. Although the derivative representation enhances the contrast of this feature, the same secondary Dirac point is also present in the raw $\sigma_{xx}$ data (Supplementary Fig. S9). For device M1, this feature appears at $n_s \approx 2.52 \times 10^{12}$ cm$^{-2}$, corresponding to an hBN/graphene alignment angle of $\theta_{hBN/graphene} \approx 0.14°$. The extraction procedure for $\theta_{hBN/graphene}$ in devices M1 and M2 is detailed in Supplementary Note 2.

At this density scale, additional curved structures emerge in the TBG spectrum that run parallel to the primary TBG branches and form satellite-like features. As indicated by the orange arrows in Fig. 3, these features appear on both the electron- and hole-doped sides and remain visible at $B = 0$ and $B = 1$ T.

Using the layer-capacitance model constrained by the extracted $C_{int}$, we convert ($n_{tot}$, $D/\varepsilon_0$) into the carrier densities of the two subsystems, $n_{graphene}$ and $n_{TBG}$. Along the curved features, $n_{TBG}$ varies only weakly, indicating that changes in $n_{tot}$ are primarily accommodated by the moiré-modulated monolayer. By identifying the total density at which the carrier density of the top monolayer crosses zero, the density positions of these curved features can be inferred. The extracted value, approximately $2.5 \times 10^{12}$ cm$^{-2}$, matches the density of the second Dirac point associated with the hBN/graphene moiré in the aligned monolayer. As the Landau-level index of the monolayer increases, additional charge is accommodated within its Landau levels, leading to a corresponding shift of the curved trajectories in total density.

These curved features are unlikely to originate from a second Dirac point in either the middle or bottom graphene layers, or from a moiré superlattice formed between the bottom monolayer graphene and the lower hBN. The middle and bottom graphene layers are twisted by $\theta_{middle/bottom} \approx 5°$; in this large-angle incommensurate regime, the corresponding moiré wavelength is short ($\lambda_m < 3$ nm), placing any associated miniband structure at energies well beyond the low-energy transport window considered here. In addition, the bottom monolayer is misaligned from the lower hBN by approximately 10°. At such an angle, the expected moiré wavelength is $\lambda_m \approx 1.4$ nm, which would correspond to superlattice minibands at energies on the order of 1 eV,[43] far above the carrier densities accessed in our measurements. This scale separation is consistent with the absence of moiré-related features associated with the bottom graphene/hBN interface within the experimental density range.

The curved features observed in the TBG spectrum are consistent with a weak moiré-periodic electrostatic modulation originating from the hBN/graphene superlattice of the aligned top monolayer, as illustrated schematically in Fig. 3c. Within our electrostatic model, the hBN/graphene moiré potential modulates the local Dirac-point energy of the top graphene and thereby generates a carrier-density modulation with the same moiré periodicity. This periodic density modulation produces a finite-wavevector Hartree potential that, although attenuated through the trilayer stack, still reaches the middle–bottom subsystem.

Within the energy range considered here, the remaining large-angle TBG retains an essentially monolayer-like Dirac electronic structure. Although the large twist angle strongly suppresses low-energy coherent interlayer hybridization, the electrostatic transfer of a long-wavelength potential is not subject to the same momentum-matching constraint. A moiré-periodic electrostatic perturbation can therefore reach the lower subsystem even when coherent tunneling between the top graphene and the lower layers is strongly suppressed.

When the attenuation of the periodic potential through the trilayer stack and Thomas–Fermi screening by both the top graphene and the lower TBG subsystem are included, our electrostatic analysis indicates that a characteristic moiré modulation of 20–60 meV acting directly on the top graphene is reduced to approximately 0.5–5.7 meV upon reaching the middle–bottom block, over the range of top-layer carrier densities sampled along the satellite trajectory. This model-dependent, order-of-magnitude amplitude is far weaker than the direct moiré potential in the top graphene, but sufficient to produce a weak periodic modulation of the lower-subsystem response. The detailed derivation and parameter dependence are provided in Supplementary Note 3.

This interpretation does not require a significant low-energy miniband reconstruction associated with the intrinsic large-angle moiré of the middle–bottom block. Rather, the data are consistent with a scenario in which the lower subsystem senses a weak periodic electrostatic potential generated by the moiré-induced density modulation of the top graphene. Within this picture, the density-locked features appear as weak satellite-like branches running parallel to the primary TBG spectrum, qualitatively consistent with partial electrostatic screening through the trilayer stack.

Furthermore, devices with larger twist angles between the top hBN and the top graphene ($\theta_{hBN/graphene}$) were fabricated to examine the angle dependence of the satellite features. Devices M2 and M3 were prepared with target alignment angles of approximately 1° and 2°, respectively. Figures 4b and 4c present derivative maps of $R_{xx}$ in the ($n_{tot}$, $D/\varepsilon_0$) plane at $B = 1$ T for these devices. Based on their alignment angles, the secondary Dirac points of the top monolayer graphene are expected at $n_s \approx 4.8 \times 10^{12}$ cm$^{-2}$ for M2 and $n_s \approx 1.2 \times 10^{13}$ cm$^{-2}$ for M3.

In device M2, the satellite features remain within the accessible measurement window. By subtracting the carrier density accommodated in the Landau levels of the top monolayer graphene from the total density, we estimate that these features appear when the TBG subsystem density is approximately $4.96 \times 10^{12}$ cm$^{-2}$ (Fig. 4b). This density corresponds to an alignment angle of about 1.03°, consistent with the target twist angle of 1°. The same density position is observed at $B = 2$ T (Supplementary Fig. S4), indicating insensitivity to magnetic field within the explored range.

By contrast, in device M3, the larger twist angle shifts the expected secondary Dirac point beyond the accessible gate-voltage window, and no corresponding satellite features are resolved. The transport response of M3 nevertheless continues to show separation into a monolayer graphene and a TBG subsystem, consistent with the large-angle decoupled regime. Upon increasing $\theta_{hBN/graphene}$ further, the system evolves toward the three-decoupled-monolayer behavior observed in device H1 (Fig. 1a). The angle-dependent evolution across devices M1–M3 is summarized in Table 1 and supports an interpretation in which the satellite features track the density scale set by the hBN/graphene moiré.

**CONCLUSIONS**

In this work, we investigated charge transport in large-angle twisted trilayer graphene, where interlayer hybridization is strongly suppressed within the low-energy regime. In the absence of hBN alignment, the system exhibits three independent Landau-level responses associated with the top, middle, and bottom graphene layers. When only the top monolayer graphene is aligned with hBN, the transport response separates into a moiré-modulated monolayer and a remaining twisted bilayer subsystem. Although the structural moiré is confined to the top interface, density scales set by the hBN/graphene moiré are reflected in the transport response of the underlying TBG block.

These results are consistent with a periodic electrostatic potential generated by the moiré-induced carrier-density modulation of the aligned top graphene and transmitted through the trilayer stack. Unlike hBN-aligned rhombohedral graphene, where moiré penetration is closely tied to strong interlayer tunneling and band hybridization, the present large-angle TTG remains weakly hybridized in the relevant energy range. The persistence of satellite-like features in this regime therefore supports an electrostatic origin, although it does not constitute direct evidence for a strongly reconstructed moiré miniband in the lower subsystem.

The trilayer geometry further shows that the modulation can extend beyond the layer directly adjacent to the moiré interface and reach both lower graphene layers. However, global transport does not directly resolve its layer-by-layer spatial profile. Local probes, layer-selective spectroscopy, or independently contacted layers would provide a more direct test. Large-angle twisted bilayer graphene with only one layer aligned to hBN may also offer a promising platform, since the shorter separation is expected to reduce vertical attenuation and could produce more pronounced satellite-like features.

## METHODS

### Device fabrication

The hBN/graphite/hBN/twisted trilayer graphene/hBN/graphite (top to bottom) heterostructures were assembled using a dry pick-up technique with an Elvacite stamp. The stack was subsequently patterned into a Hall bar geometry by electron-beam lithography followed by reactive ion etching in a $CF_4/O_2$ plasma (40/4 sccm, 40 W). Metal electrodes (Cr/Au, 5 nm/100 nm) were deposited by electron-beam evaporation. Twisted trilayer graphene was prepared using either a sequential tear-and-stack method or an AFM nanolithography approach. In the latter case, three graphene sections were defined from a single large flake and subsequently stacked with controlled relative twist angles to form the trilayer structure.[1] For devices M1–M3, which host a hBN/graphene superlattice, the crystallographic alignment between the top monolayer graphene and the adjacent hBN was achieved by aligning their straight edges during the transfer process.

### Transport measurement

Transport measurements were performed using a low-frequency lock-in technique with an AC current of 100 nA at 13.333 Hz in a $^4$He cryostat with a base temperature of 1.5 K. During measurements, we applied ±70 V to reduce contact resistance in areas not covered by the graphite gates.

## ASSOCIATED CONTENT

The preprint version of this work is available on arXiv: Kim, D.; Choe, J.; Taniguchi, T.; Watanabe, K.; Cho, G. Y.; Kim, Y. Remote Moiré Modulation of Decoupled Dirac Subsystems in Twisted Trilayer Graphene. 2026, 2604.12614. arXiv. https://arxiv.org/abs/2604.12614 (accessed September 22, 2026).

### Data Availability Statement

The data supporting the findings of this study are available from the corresponding author upon reasonable request.

### Supporting Information

The Supporting Information is available free of charge at https://pubs.acs.org.

Device information for all six devices (Table S1); additional transport data of large-angle alternating and helical twisted trilayer graphene without hBN alignment (devices A, H1, and H2); transport data of devices M2 and M3; interlayer capacitance analysis and estimation of the hBN/graphene twist angle for device M1; broken-symmetry integer quantum Hall states of the top graphene; satellite features induced by remote moiré modulation; quantitative electrostatic analysis of the remote moiré potential, including the moiré-induced density modulation, finite-wavevector Hartree transfer, and Thomas–Fermi screening of both subsystems.

## AUTHOR INFORMATION

### Corresponding Author

Youngwook Kim − Department of Physics and Chemistry, Daegu Gyeongbuk Institute of Science and Technology (DGIST), Daegu 42988, Republic of Korea; orcid.org/0000-0001-9544-5691; Email: y.kim@dgist.ac.kr

### Authors

Dohun Kim − Department of Physics and Chemistry, Daegu Gyeongbuk Institute of Science and Technology (DGIST), Daegu 42988, Republic of Korea

Junsik Choe − Department of Physics and Chemistry, Daegu Gyeongbuk Institute of Science and Technology (DGIST), Daegu 42988, Republic of Korea

Minseo Cho − Department of Physics and Chemistry, Daegu Gyeongbuk Institute of Science and Technology (DGIST), Daegu 42988, Republic of Korea

Takashi Taniguchi − Research Center for Materials Nanoarchitectonics, National Institute for Materials Science, Tsukuba 305-0044, Japan

Kenji Watanabe − Research Center for Electronic and Optical Materials, National Institute for Materials Science, Tsukuba 305-0044, Japan

Gil Young Cho − Department of Physics, Korea Advanced Institute of Science and Technology, Daejeon 34141, Republic of Korea; Center for Artificial Low Dimensional Electronic Systems, Institute for Basic Science, Pohang 37673, Republic of Korea

### Author Contributions

D. K., J. C., and Y. K. conceived the project. D. K., J. C., and M. C. carried out the device fabrication and performed the low-temperature measurements with Y. K. G. Y. C. provided theoretical support. T. T. and K. W. synthesized the hBN crystals. All authors contributed to the manuscript writing. †D. K. and J. C. contributed equally to this work.

**Notes**

The authors declare no competing financial interest.

**ACKNOWLEDGMENTS**

The work from DGIST was supported by the National Research Foundation of Korea (NRF) (Grant No. RS-2025-00557717, RS-2023-00269616, RS-2025-02315685) and the Nano and Material Technology Development Program through the National Research Foundation of Korea (NRF) funded by Ministry of Science and ICT (No. RS-2024-00444725). We also acknowledge the partner group program of the Max Planck Society. Part of this work was supported by Global Partnership Program of Leading Universities in Quantum Science and Technology (RS-2025-02317602). G. Y. C. is financially supported by Samsung Science and Technology Foundation under Project Number SSTF-BA2401-03, the NRF of Korea (Grants No. RS-2024-00410027, RS-2023-NR119931, RS-2024-00444725, RS-2023-00256050, IRS-2025-25453111, RS-2025-08542968) funded by the Korean Government (MSIT), the Air Force Office of Scientific Research under Award No. FA23862514026, and Institute of Basic Science under project code IBS-R014-D1. K.W. and T.T. acknowledge support from the CREST (JPMJCR24A5), JST and World Premier International Research Center Initiative (WPI), MEXT, Japan.

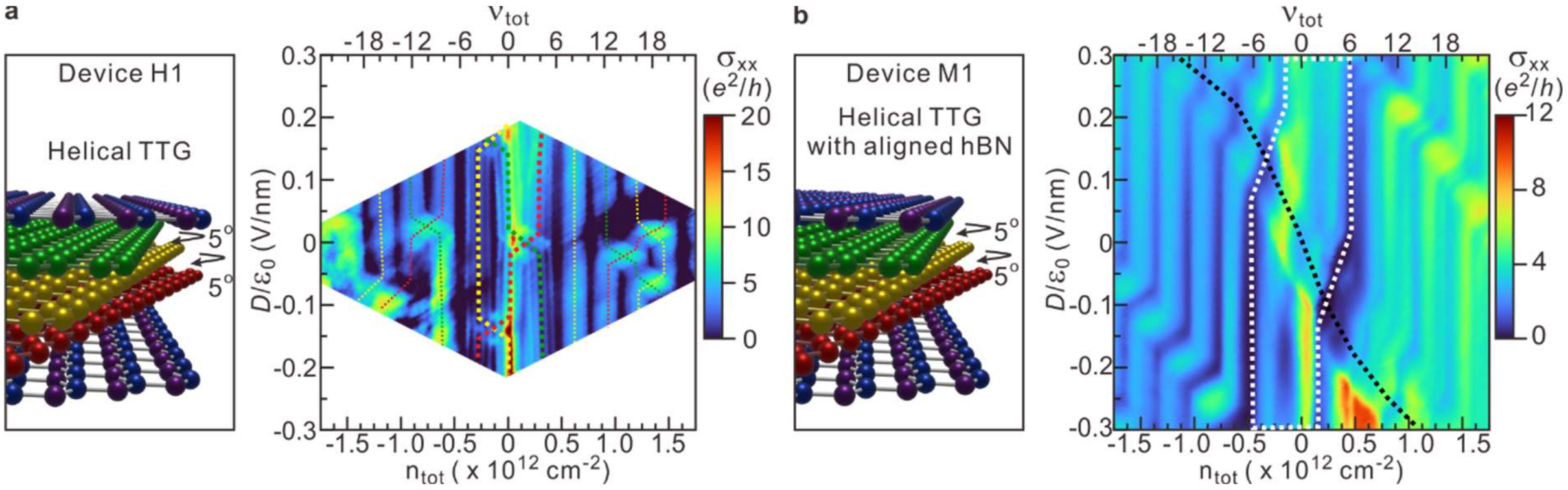

**Figure 1| Transport properties of large-angle twisted trilayer graphene** (a) Helical TTG (device H1) with $\theta_{top/middle} = \theta_{middle/bottom} \approx 5°$ (b) Helical TTG where top graphene and hBN are aligned with $\theta_{hBN/graphene}$ (device M1). The relative interlayer twist angles are $\theta_{top/middle} = \theta_{middle/bottom} \approx 5°$. The schematics shown to the left of each contour map illustrate the corresponding twisted trilayer graphene configurations. In panel a, green, yellow, and red dotted lines represent the Landau levels from top, mid, and bottom graphene, respectively. In panel b, the black and white dotted lines denote the lowest Landau levels of the top graphene and the remaining TBG, respectively. All data were recorded at $B$ = 3 T.

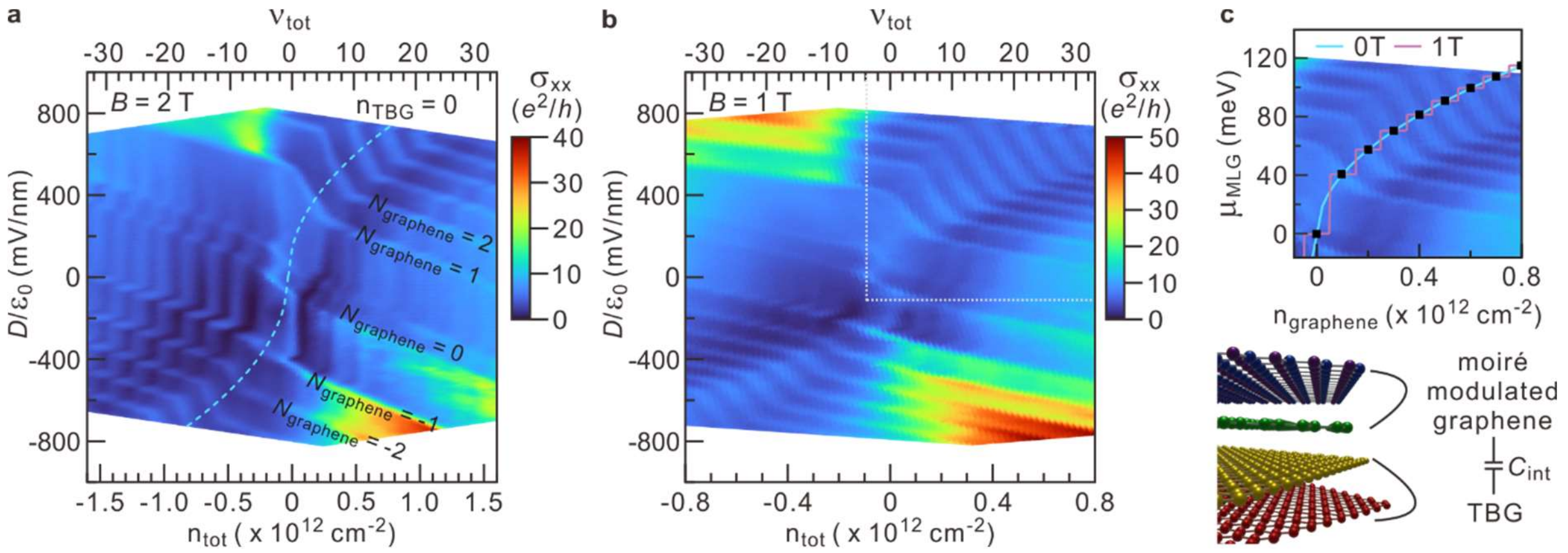

**Figure 2| Interlayer capacitance in helical twisted trilayer graphene with aligned top hBN.** (a) The contour map of $\sigma_{xx}$ at $B$ = 2 T. The numbers on diagonal features correspond to Landau level index of top graphene. Quantum Hall states with $\Delta\nu_{tot} = 4$ in vertical lines are from the layer symmetry-broken quantum Hall states from TBG subsystem. A cyan dotted line marks the charge neutrality point of TBG. (b) same as panel a but for 1T. (c) The chemical potential of top graphene as a function of graphene's density ($n_{graphene}$) at $B$ = 0 T(cyan) and 1 T(purple) calculated with $v_F \approx 1.12 \times 10^6$ m/s. The background figure is identical to white dotted regime in panel b and black dots mark the charge neutral regime of TBG subsystem satisfying $n_{tot} = n_{graphene}$. Interlayer capacitance ($C_{int}$) between top graphene and bottom TBG subsystem depicted in lower figure obtained from matching the ($V_t$, $V_b$) of black dots.

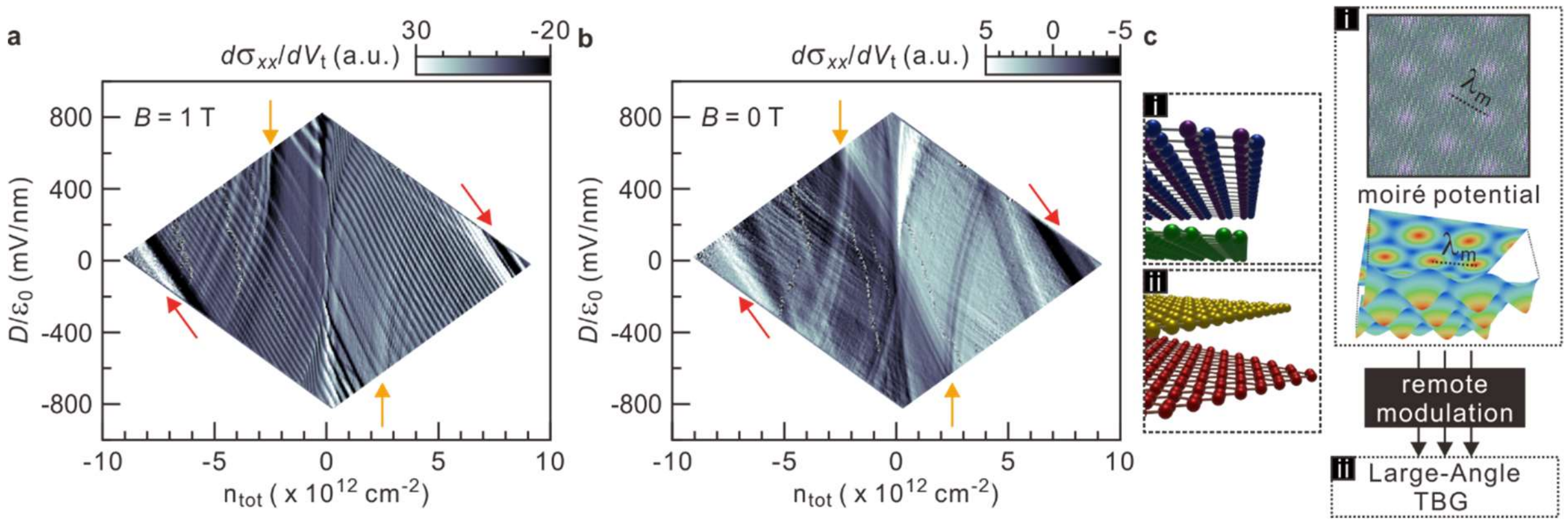

**Figure 3| Transport signatures of moiré-induced modulation in the TBG subsystem.** (a) Derivative of $\sigma_{xx}$ in ($n_{tot}$, $D/\varepsilon_0$) plane at $B$ = 1 T. (b) same as panel a but for $B$ = 0 T. Red and orange arrows highlight the second Dirac point of hBN/graphene superlattice probed in top graphene and in TBG subsystem, respectively. All data were recorded at $T$ = 1.5 K. (c) Schematic of moiré potential effect. Alignment of top graphene with hBN generates a moiré-induced density modulation in the top layer, whose periodic electrostatic potential can influence the underlying TBG subsystem.

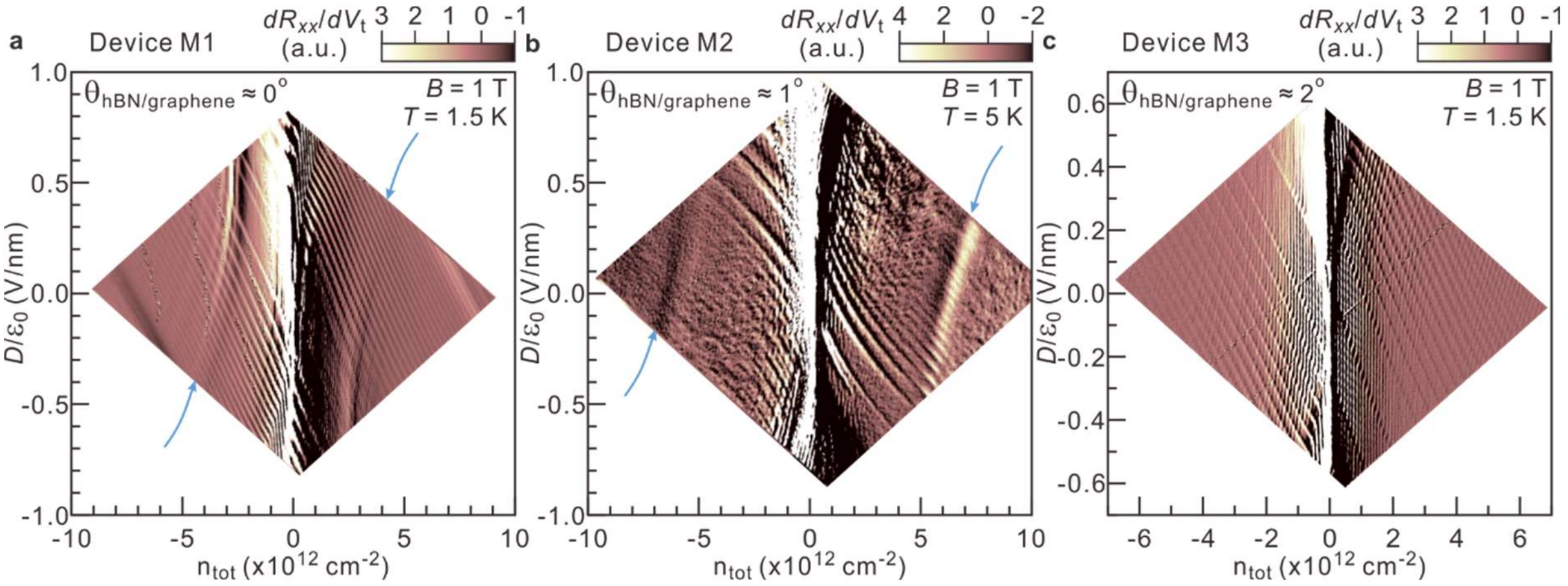

**Figure 4| Dependence of moiré-induced satellite features on the hBN/graphene alignment angle.** (a) Derivative of $R_{xx}$ represented in the $n_{tot}$ - $D/\varepsilon_0$ space for device M1 with a twist angle of $\theta_{hBN/graphene} \approx 0°$. (b, c) Same as panel (a), but for $\theta_{hBN/graphene} \approx 1°$ and 2°, respectively. Blue arrows in panel (a) and (b) indicate satellite features attributed to moiré-induced electrostatic modulation.

| | Device M1 | Device M2 | Device M3 | Device H1 | Device H2 |
|---|---|---|---|---|---|
| ($\theta_{top/middle}$, $\theta_{middle/bottom}$) | ~ (5°, 5°) | ~ (5°, 5°) | ~ (5°, 5°) | ~ (5°, 5°) | ~ (2°, 2°) |
| $\theta_{hBN/graphene}$ | ~ 0° | ~ 1° | ~ 2° | > 5° | > 5° |
| subsystems | graphene + TBG | graphene + TBG | graphene + TBG | three decoupled graphenes | three decoupled graphenes |
| 2$^{nd}$ DP signal | Observed | Not accessible | Not accessible | None | None |
| Moiré-induced Satellite peaks | Observed | Observed | Not accessible | None | None |

**Table 1| Evolution of subsystem decoupling and moiré-induced satellite features with twist angle.** Device parameters and the corresponding transport characteristics summarized for device M1, M2, M3, H1, and H2.

# Supporting Information on **"Remote Moiré Modulation of Decoupled Dirac Subsystems in Twisted Trilayer Graphene"**

Dohun Kim[1,†], Junsik Choe[1,†], Minseo Cho[1], Takashi Taniguchi[2], Kenji Watanabe[3], Gil Young Cho[4,5], and Youngwook Kim[1,*]

[1]*Department of Physics and Chemistry, Daegu Gyeongbuk Institute of Science and Technology (DGIST), Daegu 42988, Republic of Korea*

[2]*Research Center for Materials Nanoarchitectonics, National Institute for Materials Science, 1-1 Namiki, Tsukuba 305-0044, Japan*

[3]*Research Center for Electronic and Optical Materials, National Institute for Materials Science, 1-1 Namiki, Tsukuba 305-0044, Japan*

[4]*Department of Physics, Korea Advanced Institute of Science and Technology, Daejeon 34141, Republic of Korea*

[5]*Center for Artificial Low Dimensional Electronic Systems, Institute for Basic Science, Pohang 37673, Korea*

[*]E-mail: y.kim@dgist.ac.kr

## SUPPLEMENTARY NOTE 1: ADDITIONAL TRANSPORT DATA OF TWISTED TRILAYER GRAPHENE

We measured a total of six twisted trilayer graphene devices, summarized in Table S1. In devices A, H1 and H2, the graphene layer adjacent to hBN is misaligned by more than 5°, such that moiré potential effects are negligible. In this regime, the transport response is consistent with three decoupled monolayer graphene subsystems.

By contrast, devices M1–M3 have smaller twist angles between graphene and hBN, resulting in a separation into a moiré-modulated graphene subsystem and a remaining twisted bilayer graphene (TBG) subsystem.

Device M1, discussed in the main text, has $\theta_{hBN/graphene}$ close to 0°, allowing both the secondary Dirac point of the top graphene and the associated moiré-induced satellite features to be resolved within the available gate range. For device M2, with $\theta_{hBN/graphene} \approx 1°$, the moiré-induced satellite features remain detectable, whereas the secondary Dirac point lies outside the experimentally accessible gate-voltage window. In device M3, with $\theta_{hBN/graphene} \approx 2°$, neither the secondary Dirac point nor the corresponding satellite features fall within the accessible density range.

| | Device M1 | Device M2 | Device M3 | Device A | Device H1 | Device H2 |
|---|---|---|---|---|---|---|
| Gate (top, bottom) | graphite | Metal, $SiO_2$/Si | graphite | graphite | Metal, $SiO_2$/Si | Metal, $SiO_2$/Si |
| ($\theta_{top/middle}$, $\theta_{middle/bottom}$) | ~ (5°, 5°) | ~ (5°, 5°) | ~ (5°, 5°) | ~ (5°, -5°) | ~ (5°, 5°) | ~ (2°, 2°) |
| $\theta_{hBN/graphene}$ | ~ 0° | ~ 1° | ~ 2° | > 5° | > 5° | > 5° |
| subsystems | graphene + TBG | graphene + TBG | graphene + TBG | three decoupled graphene | Three decoupled graphene | Three decoupled graphene |
| 2nd DP signal | Observed | Not accessible | Not accessible | None | None | None |
| Moiré induced satellite peak | Observed | Observed | Not accessible | None | None | None |

**Table S1| Device information.**

## Helical and Alternating Twisted trilayer graphene with misaligned hBN

For both alternating and helical twisted trilayer graphene (TTG), large interlayer twist angles place the system in a regime where the three graphene layers behave as decoupled Dirac systems [2]. Their transport responses are therefore similar in this limit.

Figure S1 presents schematics and optical microscope images of an alternating TTG device (device A) and two helical TTG devices (device H1 in the main text and device H2 with smaller twist angles). Figure S2 shows transport data from device A with twist angles $\theta_{top/middle} = -\theta_{middle/bottom} = 5°$ and device H2 with $\theta_{top/middle} = \theta_{middle/bottom} = 2°$. As in Fig. 1a of the main text, the conductance map displays two diagonal branches with opposite slopes (green and red dotted lines) together with nearly vertical features (yellow dotted lines). These branches arise from Landau levels of three decoupled monolayer graphene sheets. Although device H2 has a much smaller twist angle of 2° than device H1 in the main text, interlayer tunneling remains sufficiently suppressed, resulting in well-resolved layer-specific Landau levels.

The outer graphene layers couple predominantly to the top and bottom gates, producing diagonally dispersing features. The middle layer, screened by the outer layers, is comparatively insensitive to the applied displacement field [3,4], resulting in the nearly vertical features observed in the transport maps.

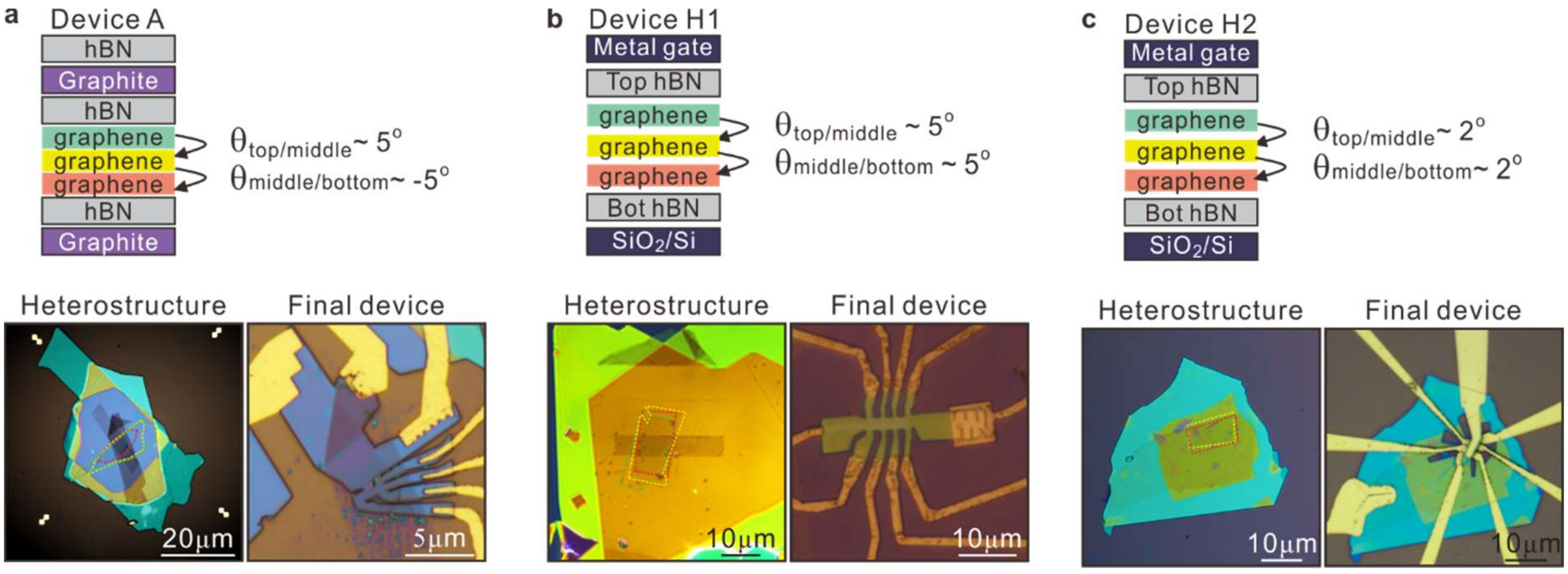

**Fig S1| Schematic and optical microscopic images of alternating and helical TTGs.** (a) Configuration and micrograph of Alternating TTG, device A. (b, c) same as panel a but for device H1 and H2. Green, yellow, red dotted lines in microscopic images correspond to top, middle, and bottom graphene layers, respectively.

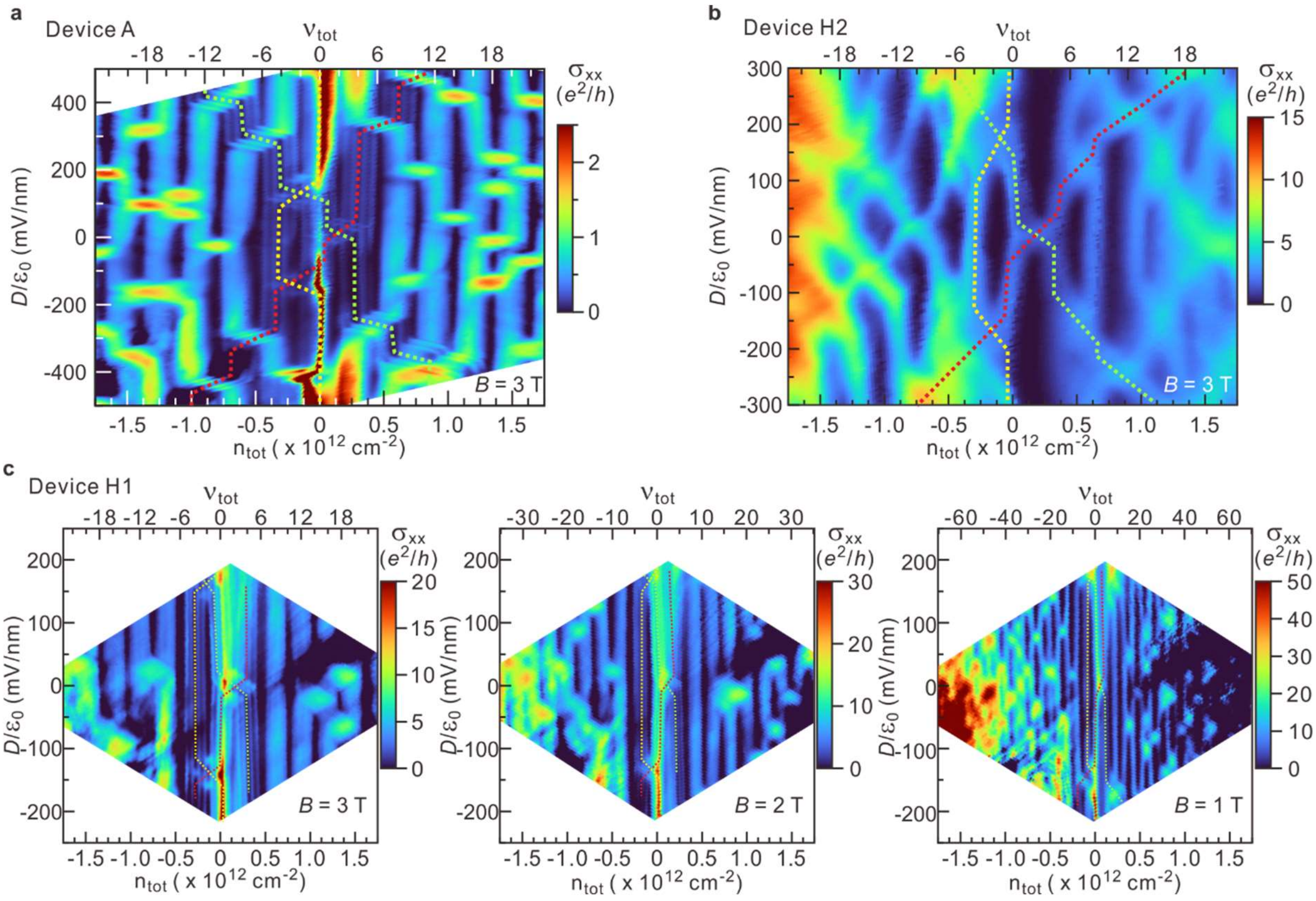

**Fig S2| Transport properties of large-angle alternating and helical twisted trilayer graphene.** (a, b) Contour map of $\sigma_{xx}$ spanned in ($n_{\text{tot}}$, $D/\varepsilon_0$) at $B$ = 3 T in device A and device H2, respectively. (c) $\sigma_{xx}$ as a function of total density and displacement field at $B$ = 3, 2, and 1 T in device H1. Green, yellow, and red dotted lines represent the lowest Landau level of top, mid, and bottom graphene.

## Twisted Trilayer graphene with aligned hBN

We fabricated and measured three helical twisted trilayer graphene devices (M1, M2, and M3) with target crystallographic alignment angles of approximately 0°, 1°, and 2° between the top hBN and the top graphene layer, as summarized in Fig. S3. In all three devices, the relative twist angles between adjacent graphene layers are approximately 5°, placing the trilayer firmly in the large-angle regime.

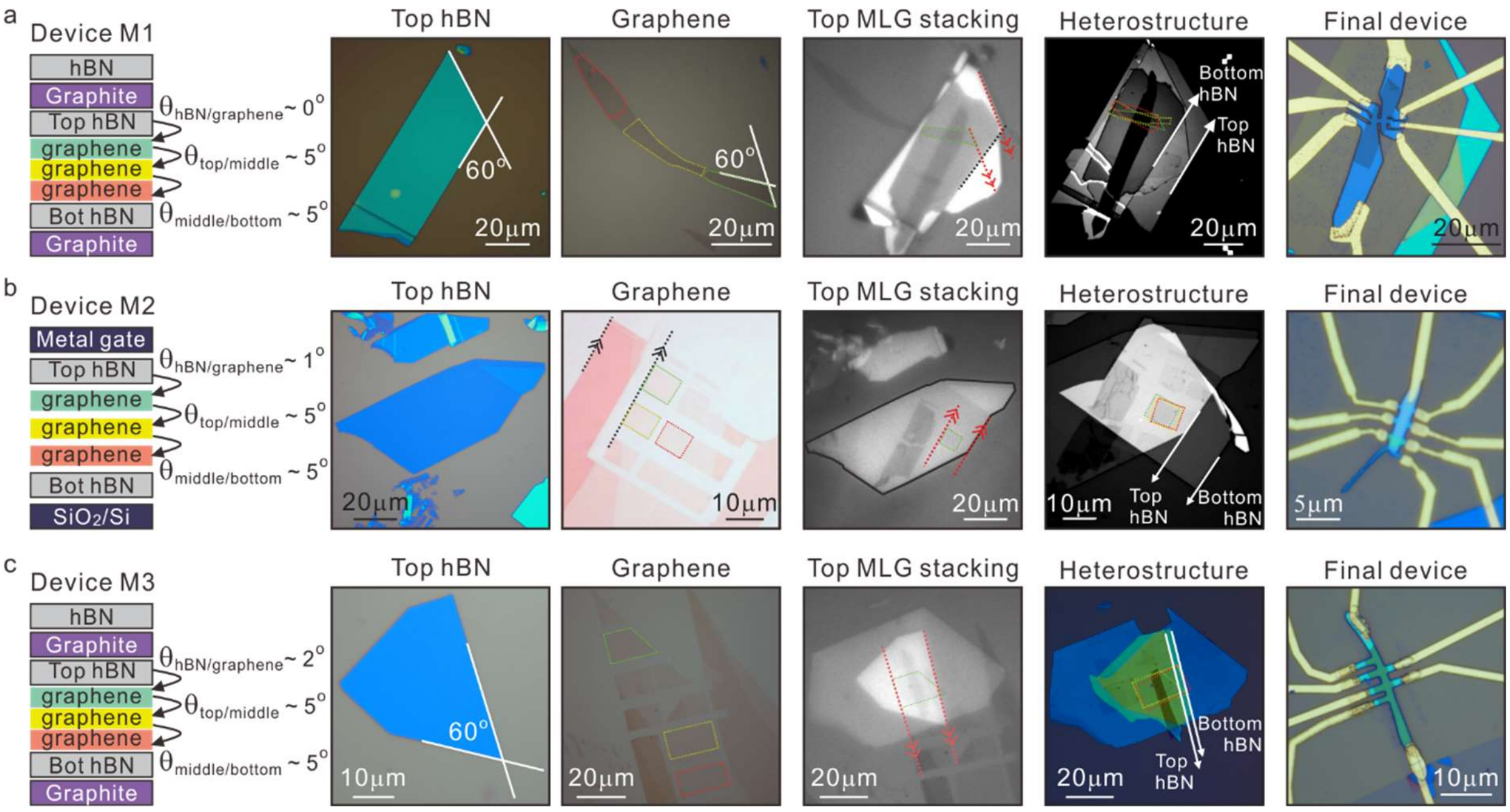

**Fig S3| Schematic and optical micrograph of device M1, M2 and M3.** (a-c) Helical TTG with aligned top hBN. Green, yellow, red dotted lines in microscopic images correspond to top, middle, and bottom graphene layers, respectively. For all devices, the crystallographic alignment between the top hBN and top monolayer graphene can be confirmed in the top monolayer graphene stacking image. In all devices, the bottom graphene and the bottom hBN are misaligned by an angle of approximately 10°.

Figure S4 presents transport data for devices M2 and M3. In both cases, the maps display nearly vertical features together with a single diagonal branch, consistent with the subsystem structure observed in device M1 (Fig. 1b in the main text). As $\theta_{hBN/graphene}$ increases (M1 < M2 < M3), both the secondary Dirac point of the top monolayer graphene and the associated moiré-induced satellite features shift to higher carrier densities.

In device M2, the moiré satellite feature remains within the accessible density window, whereas the secondary Dirac point lies beyond it. In Fig. S4a, the blue arrows mark the satellite feature. The corresponding TBG carrier density is obtained by subtracting the density associated with the quantum Hall state of the top monolayer graphene, yielding $n_{TBG} \approx 4.96 \times 10^{12}$ cm$^{-2}$. This density corresponds to an alignment angle of approximately 1.03° between the top hBN and the top graphene layer. For this angle, the secondary Dirac point of the top monolayer is expected at the $N = 52$ Landau level at $B = 1$ T, which exceeds the experimentally accessible density range.

For device M3, fabricated with a target twist angle of approximately 2°, the expected density of the secondary Dirac point is $n_s \approx 1.18 \times 10^{13}$ cm$^{-2}$, which likewise lies outside the measurement window. As a result, neither the secondary Dirac point nor the associated satellite feature is resolved in the accessible range.

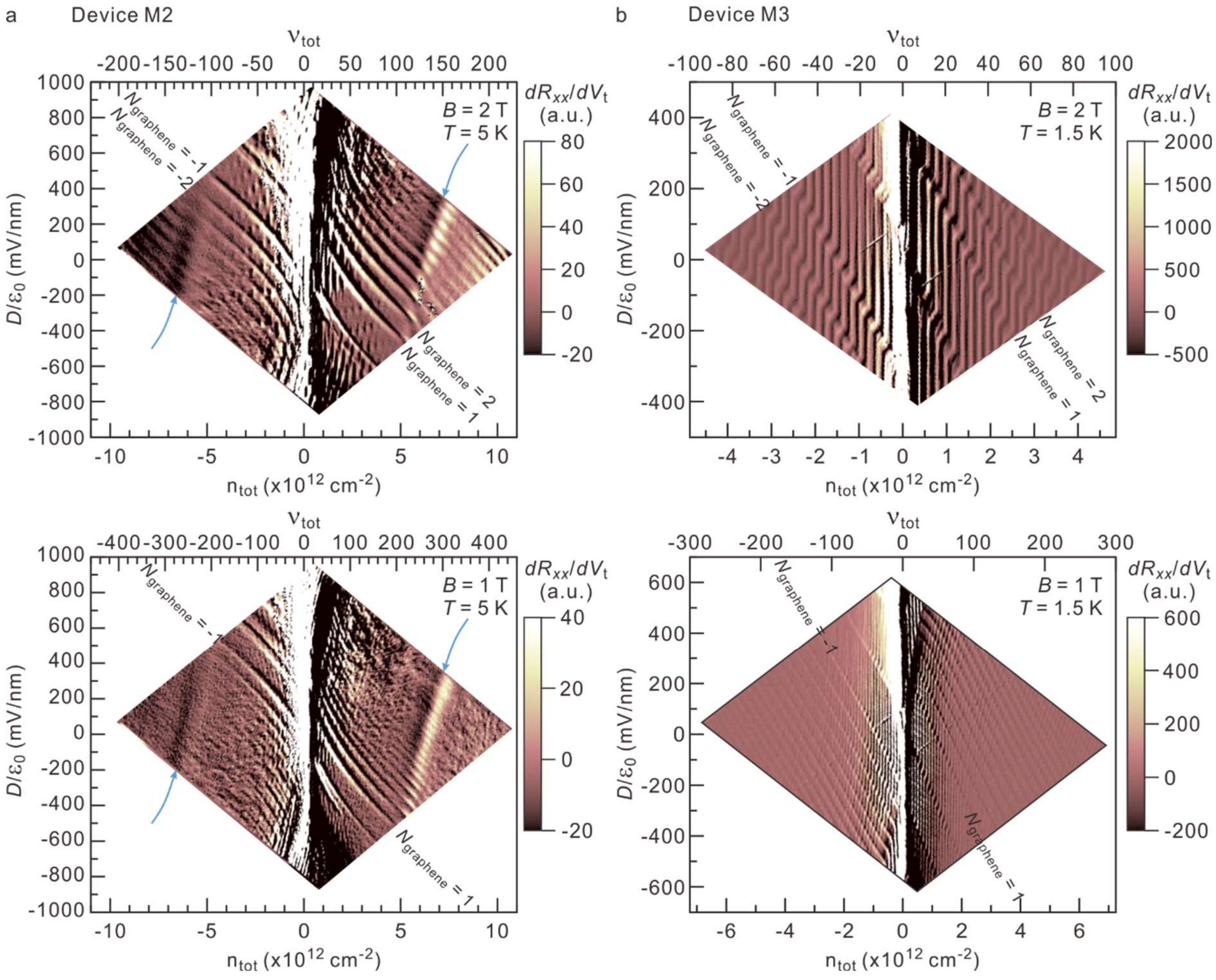

**Fig S4| Transport properties of device M2 and M3.** (a) Derivative of $R_{xx}$ in the plane of ($n_{tot}$, $D/\varepsilon_0$) for device M2 at $B$ = 1 and 2 T. Blue arrows indicate the moiré satellite signal. (b) Same as panel a but for device M3. Due to relatively large $\theta_{hBN/graphene}$, both moiré imprinting and 2$^{nd}$ Dirac point in graphene features are not observed. Data of devices M2 and M3 were obtained at $T$ = 5 and 1.5 K, respectively.

## SUPPLEMENTARY NOTE 2: DETAILED ANALYSIS OF DEVICE M1

### Interlayer Capacitance Calculation

In a TTG system hosting a hBN/graphene moiré superlattice on the outermost graphene layer, the system can be effectively decomposed into a moiré-modulated graphene subsystem and a remaining TBG subsystem. In multilayer graphene systems, electrostatic screening modifies the electric fields generated by the top and bottom gates. When the moiré graphene subsystem and the TBG subsystem are in equilibrium, their electrochemical potentials are equal. Under this condition, the interlayer capacitance between the two subsystems can be estimated using the capacitance model described below [5,6].

Equation (1) is derived for a configuration consisting of a top graphene and a bottom TBG:

$$
\begin{aligned}
en_{graphene} &= C_t\left(V_t - \frac{\mu_{graphene}}{e}\right) - C_{int}\left(\frac{\mu_{graphene}}{e} - \frac{\mu_{TBG}}{e}\right) \\
en_{TBG} &= C_b\left(V_b - \frac{\mu_{TBG}}{e}\right) + C_{int}\left(\frac{\mu_{graphene}}{e} - \frac{\mu_{TBG}}{e}\right)
\end{aligned}
\tag{1}
$$

Here, $C_t$ and $C_b$ denote the geometric capacitances of the top and bottom hBN layers, respectively, and $C_{int}$ is the interlayer capacitance between the top graphene and the bottom TBG. $V_t$ and $V_b$ are the applied gate voltages, and $\mu_{graphene}$ and $\mu_{TBG}$ are the corresponding chemical potentials.

Setting $n_{TBG} = 0$ and $\mu_{TBG} = 0$ in Eq. (1), corresponding to the TBG charge neutrality point (CNP), we obtain

$$
\begin{aligned}
\mu_{MLG} &= -\frac{eC_bV_b}{C_{int}} \\
n_{MLG} &= \frac{C_tV_t}{e} + \frac{(C_t + C_{int})C_bV_b}{eC_{int}}
\end{aligned}
\tag{2}
$$

If the system were instead composed of a top TBG and a bottom graphene, the roles of $C_t$ and $C_b$, as well as $V_t$ and $V_b$, would be interchanged in Eq. (2). Evaluating the data under this assumption yields a negative value of the interlayer capacitance $C_{int}$, which is not physically meaningful. Therefore, only the configuration consisting of a top graphene and a bottom TBG is consistent with our observations.

At zero magnetic field, the chemical potential of monolayer graphene is given by $\mu_{graphene} = sgn(n_{graphene})\hbar v_F(\pi|n_{graphene}|)^{1/2}$. In the experiment, the TBG charge neutrality point is identified at a small magnetic field of $B$ = 1 T as the midpoint between the two kink lines marked in Fig. S5(a). At finite magnetic field, Landau quantization leads to a staircase-like evolution of $\mu_{graphene}$ as a function of carrier density. Nevertheless, the overall envelope of the chemical potential remains well described by the zero-field Dirac dispersion, as demonstrated in previous studies [6]. We therefore use the above relation to estimate $\mu_{graphene}$ from the monolayer graphene density corresponding to given Landau level indices $N_{graphene}$.

Throughout this analysis, we adopt a bare graphene Fermi velocity $v_F = 1.12 \times 10^6$ m/s [6]. While the formation of a hBN/graphene moiré superlattice can, in principle, renormalize the Fermi velocity, such effects lead to only a few-percent correction in the low-energy regime relevant here, rendering this approximation valid [7].

The interlayer capacitance $C_{int}$ is extracted by identifying the TBG charge neutrality point from the contour maps of $\sigma_{xx}$ in the ($V_t$, $V_b$) plane (black dots in Fig. S5). For device M1, the geometric capacitances of the top and bottom hBN layers are taken as $C_t = 5.05 \times 10^{-2}$ μF/cm$^2$ and $C_b = 5.30 \times 10^{-2}$ μF/cm$^2$.

Table S2 summarizes the values of $C_{int}$ obtained at different TBG neutrality points corresponding to specific Landau level indices $N_{graphene}$. Averaging over these data points yields $C_{int} = 6.12 \pm 0.14$ μF/cm$^2$.

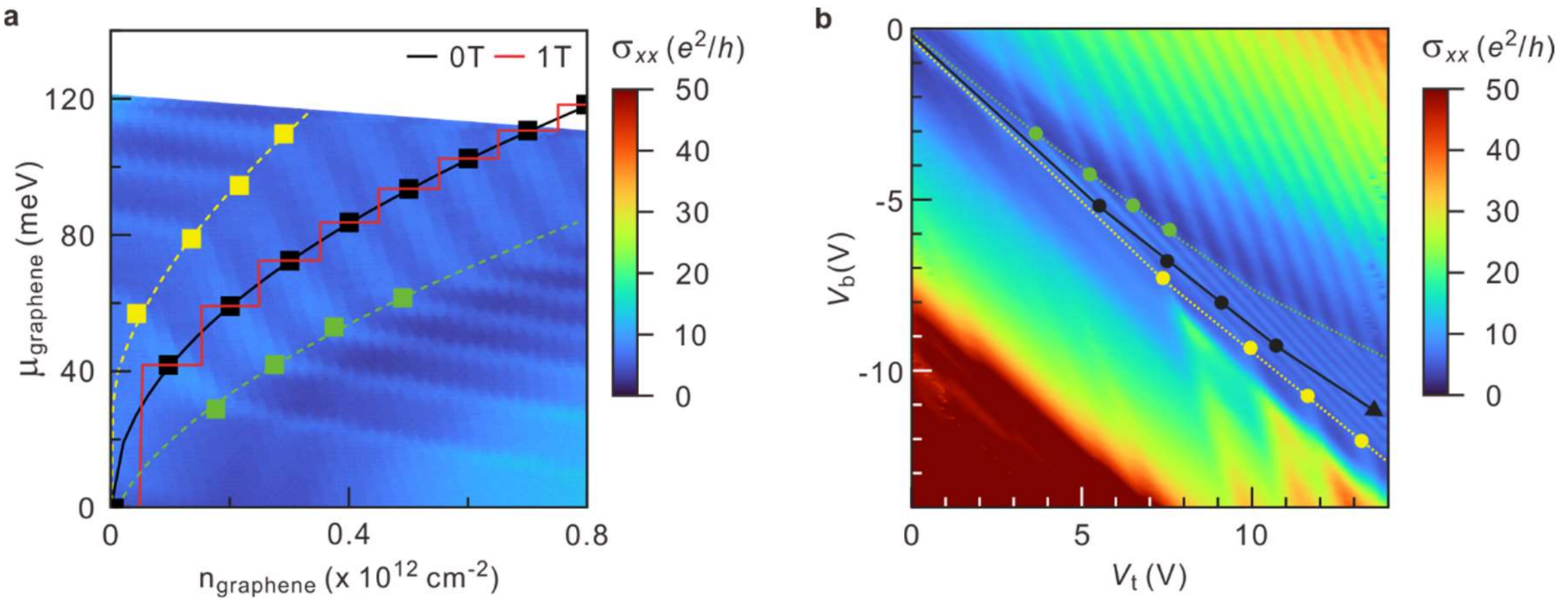

**Fig S5| Contour maps of $\sigma_{xx}$ measured at $B$ = 1T.** (a) Chemical potential of top graphene as a function of $n_{graphene}$, overlaid with the contour map of $\sigma_{xx}$ plotted in the ($n_{tot}$, $D/\varepsilon_0$) plane. (b) $\sigma_{xx}$ data in ($V_t$, $V_b$) map. The green and yellow dotted lines indicate the locations where the slope changes abruptly along the Landau level lines of the top graphene. The midpoints between the corresponding green and yellow dots are marked by black dots, which denote the charge neutrality point of the TBG subsystem.

| $N_{graphene}$ | $\mu_{graphene}$ (J) | $V_b$ at CNP of TBG (V) | $C_{int}$ (μF/cm$^2$) |
|---|---|---|---|
| 1 | $6.509 \times 10^{-21}$ | -4.935 | 6.43 |
| 2 | $9.204 \times 10^{-21}$ | -6.79 | 6.26 |
| 3 | $1.127 \times 10^{-20}$ | -7.91 | 5.95 |
| 4 | $1.302 \times 10^{-20}$ | -8.96 | 5.84 |

**Table S2| Estimated $C_{int}$ at $B$ = 1 T contour map.**

## Estimation of Twist Angle in the hBN/graphene Superlattice

In twisted trilayer graphene, there is no insulating spacer separating the graphene layers. Therefore, transport measurements reflect the combined response of the entire multilayer stack rather than that of an individual layer. When probing features originating from the top graphene, such as the second Dirac point induced by the hBN/graphene moiré superlattice, the measured signal inevitably contains background contributions from the remaining graphene layers.

For this reason, it is difficult to determine the carrier density corresponding to the second Dirac point using only zero-magnetic-field data. For example, simply dividing the total carrier density at which the second Dirac point appears by three does not provide an accurate estimate. Each graphene layer experiences a slightly different electrostatic potential shift, and in device M1 the top graphene additionally forms a moiré superlattice with hBN, which substantially modifies its band structure.

This ambiguity can be resolved by applying a small magnetic field ($B \approx 1$ T). Under this condition, Landau levels of the top graphene become distinguishable. By identifying the Landau level index of the top graphene at the position of the second Dirac point, the corresponding carrier density ($n_s$) can be determined.

Figure S6 shows the derivative of $\sigma_{xx}$ in the ($V_t$, $V_b$) plane measured at $B$ = 1 T, together with line cuts of $R_{xx}$ as a function of $V_t$ taken at $V_b$ = 12.53 V. Although the transport signals from the top graphene and the other two graphene layers overlap, the enhanced resistance associated with the second Dirac point remains identifiable above the background contributions, allowing reliable analysis. In Fig. S6, the second Dirac point is located at Landau level index $N_{graphene}$ = 26. The corresponding filling factor is $\nu_{graphene}$ = 104, which yields a carrier density $n_s = 2.52 \times 10^{12}$ cm$^{-2}$.

Using this value of ns, the moiré wavelength ($\lambda_m$) and the twist angle ($\theta_{hBN/graphene}$) can be obtained from

$$n_s = \frac{8}{\sqrt{3}\lambda_m^2} = \frac{8}{\sqrt{3}a^2}\left(\theta_{hBN/graphene}^2 + \delta^2\right) \quad (3)$$

where a is the lattice constant of graphene and δ = 1.8% represents the lattice mismatch between graphene and hBN [7–9]. For device M1, this analysis gives a moiré wavelength $\lambda_m \approx 13.5$ nm and a twist angle $\theta_{hBN/graphene} \approx 0.14°$.

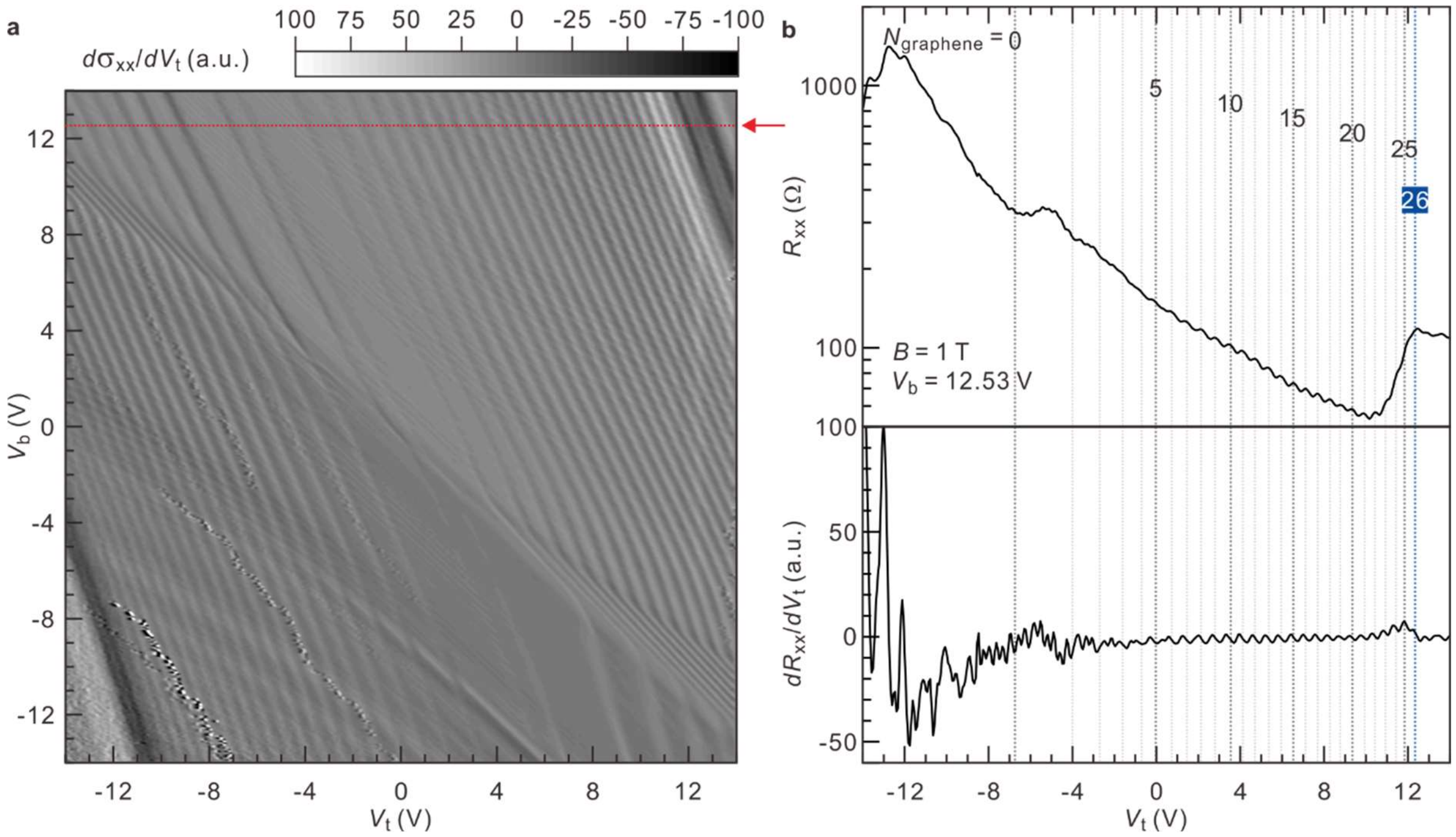

**Fig S6| Estimation of $n_s$ for device M1.** (a) $d\sigma_{xx}/dV_t$ contour map at $B$ = 1 T. (b) $R_{xx}$ (upper) and derivative of $R_{xx}$ (lower) line data as a function of $V_t$ at $V_b$ = 12.53 V indicated by red lines in panel a. The numbers in panel b represent the Landau level indices of top graphene. A blue line highlights the second Dirac point of top graphene.

## Broken symmetry integer quantum Hall states of top graphene

Figure S7 shows that the valley symmetry breaking observed in the lowest Landau level ($N_{graphene}$ = 0), which is visible at $B$ = 2 T and 1 T in Fig. 2 of the main text, persists down to zero magnetic field. The persistence of this splitting at $B$ = 0 indicates that the valley symmetry breaking is associated with the formation of the hBN/graphene moiré superlattice, which breaks the sublattice (inversion) symmetry of graphene.

At higher magnetic fields, the top monolayer graphene exhibits quantum Hall states in which both spin and valley degeneracies are lifted. Figure S8 presents contour maps of $\sigma_{xx}$ in the ($\nu_{tot}$, $D/\varepsilon_0$) plane measured at $B$ = 6 T and 9 T for device M1. In the lower-field data shown in Figs. 1b, 2, and 3 of the main text, the hBN/graphene moiré superlattice breaks inversion symmetry and opens a gap at the Dirac point. As a result, the $N_{graphene}$ = 0 Landau level primarily displays valley splitting, while the spin degeneracy remains unresolved.

At higher magnetic fields, the spin degeneracy is also lifted, leading to four resolved sublevels within the lowest Landau level. Fully symmetry-resolved quantum Hall states, with both spin and valley degeneracies lifted, are likewise observed in the $N_{graphene}$ = −1 Landau level.

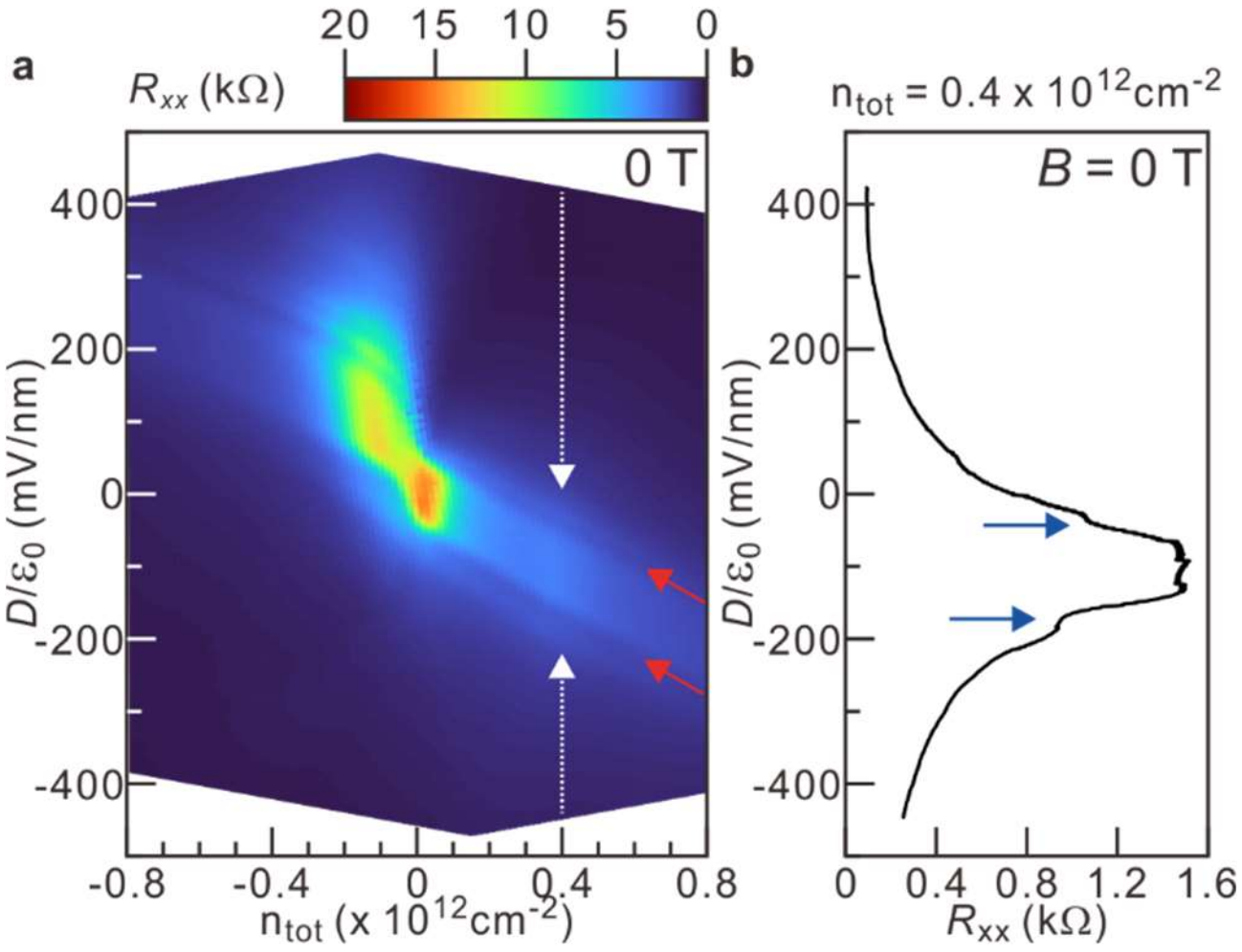

**Figure S7| Valley symmetry-broken quantum Hall states induced by hBN/graphene superlattice.** (a) Contour map of $R_{xx}$ in ($n_{tot}$, $D/\varepsilon_0$) plane at $B$ = 0 T. The red arrows in contour map indicate the valley symmetry-broken quantum Hall states. (b) The line trace of $R_{xx}$ for $B$ = 0 T as a function of displacement field at white line in panel a. The shallow dips marked by blue arrows represent the valley symmetry-broken states.

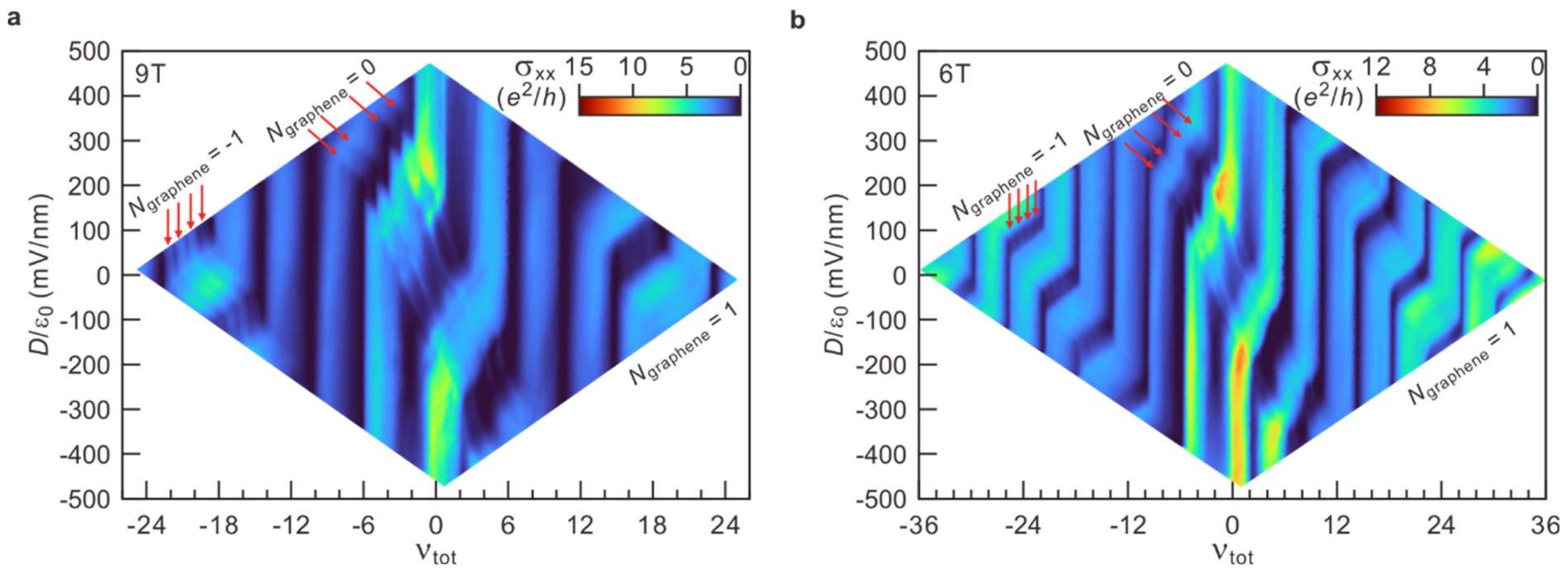

**Fig S8| Broken symmetry integer quantum Hall states of top graphene for device M1.** (a) Contour map of $\sigma_{xx}$ in ($\nu_{tot}$, $D/\varepsilon_0$) plane obtained at $B$ = 9T. (b) same as b but for $B$ = 6 T. The red arrows mark the Landau levels where both spin and valley degeneracies are fully lifted. All data were recorded at $T$ = 1.5 K.

## Satellite features induced by remote moiré modulation

Figure S9 presents contour maps of $R_{xx}$, $R_{xx}^{-1}$, and derivatives of $R_{xx}^{-1}$ and $R_{xy}$ measured at $B$ = 0, 1, and 2 T. The curved lines marked by blue arrows correspond to enhanced resistance relative to the surrounding regions. These features are more clearly resolved in $R_{xx}^{-1}$ and its derivatives, as well as in the derivative maps of $R_{xy}$. Along these curved trajectories, the carrier density of the TBG subsystem remains approximately constant, while the density of the top monolayer graphene varies. This behavior is confirmed by the discrete shifts of the feature observed at $B$ = 1 T and 2 T when the Landau level index of the top graphene changes. The discontinuous shifts indicate that the curved feature follows the density evolution of the top graphene rather than that of the TBG subsystem.

Given the geometric configuration of device M1, these curved features are unlikely to originate from a moiré superlattice formed between graphene layers or between the bottom graphene layer and the bottom hBN. Instead, they are consistent with a periodic electrostatic modulation associated with the top graphene–hBN alignment, which is sensed by the underlying TBG subsystem. The red dashed line in Fig. S9 marks the charge neutrality point of the top graphene. At the intersection between this line and the curved feature (blue arrows), the total carrier density equals the carrier density of the TBG subsystem. The curved features occur at approximately $n_{TBG} = \pm 2.5 \times 10^{12}$ cm$^{-2}$, matching the density scale of the hBN/graphene moiré. This agreement is consistent with a remote moiré-modulation mechanism. Finally, the curved features are more pronounced on the hole-doped side. The electron–hole asymmetric visibility of the satellite-like feature most plausibly reflects the intrinsic electron–hole asymmetry of the graphene/hBN moiré response. Previous studies have consistently reported that secondary-Dirac-point signatures in graphene/hBN superlattices are more pronounced on the hole side than on the electron side [7,8,10,11]. In our model, the hBN/top-graphene moiré first creates a density modulation in the top graphene, and this source modulation generates the finite-$q$ Hartree potential transferred to the lower subsystem [see Supplementary Note 3]; an electron–hole asymmetric source modulation therefore translates directly into an asymmetric transferred amplitude. Density-dependent screening and parallel-channel transport may further modulate the visibility. The simple scalar estimate presented here is not intended to reproduce this asymmetry quantitatively.

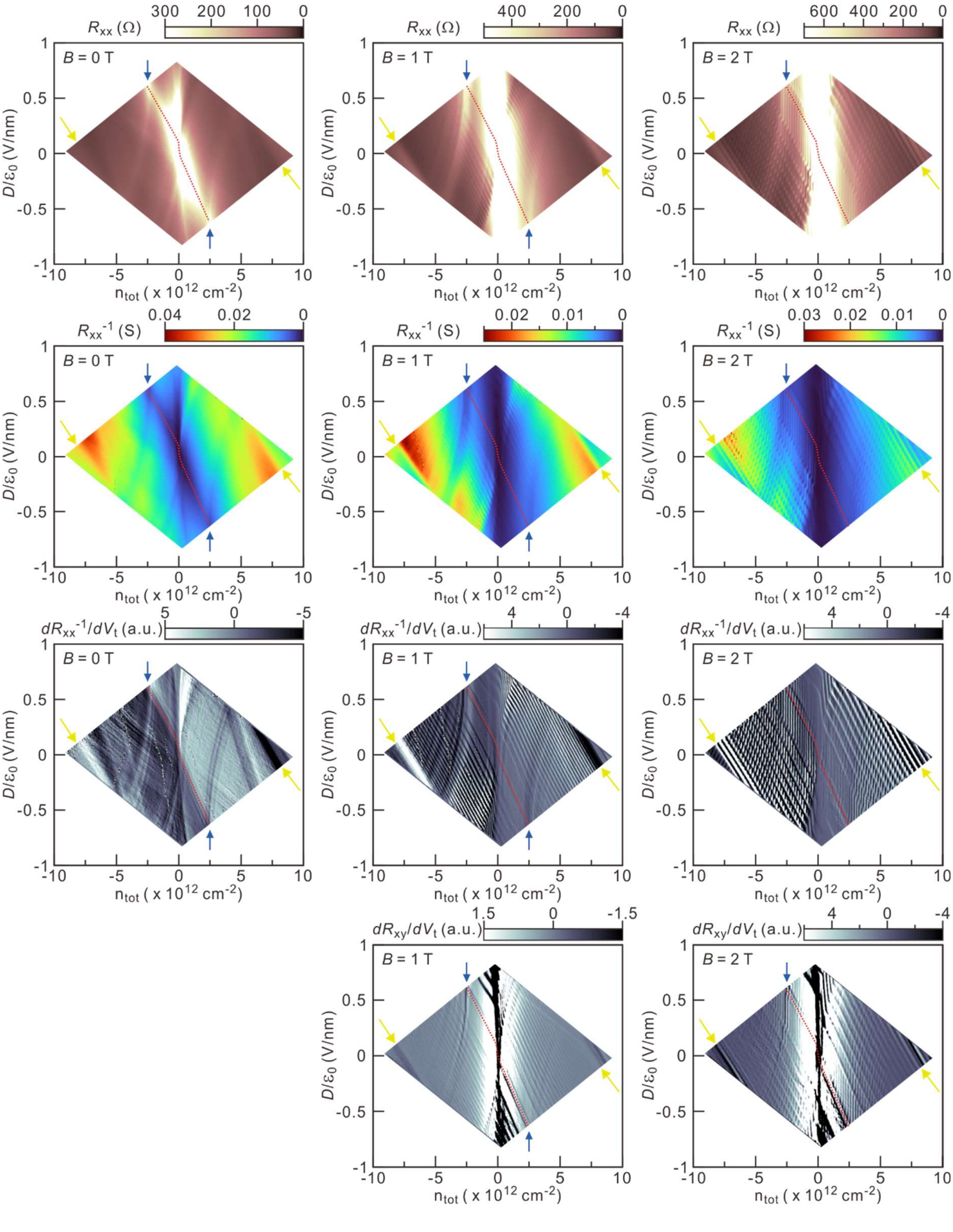

**Fig S9| Moiré satellite signal at 0, 1, and 2 T for device M1.** $R_{xx}$, $R_{xx}^{-1}$, derivatives of $R_{xx}^{-1}$ and $R_{xy}$ at various magnetic field. The yellow, blue arrows denote the 2nd DP of the top graphene and the moiré satellite signal experienced by the TBG, respectively. The red dotted lines represent the charge neutrality point of top graphene. All data were recorded at $T$ = 1.5 K.

## SUPPLEMENTARY NOTE 3: QUANTITATIVE ELECTROSTATIC ANALYSIS OF THE REMOTE MOIRÉ POTENTIAL

Our central picture is that the hBN/graphene moiré potential first modulates the local Dirac-point energy of the top graphene. This produces not only an average charge redistribution but also a spatially periodic density modulation at the moiré wavevector. The corresponding finite-$q$ Hartree potential propagates through the trilayer thickness and acts as a weak periodic perturbation on the lower subsystem.

### Moiré-Induced Density Modulation in the hBN-Aligned Top Graphene

For the purpose of estimating the electrostatic transfer of the remote moiré potential, we keep only the scalar first-harmonic component of the hBN/graphene moiré perturbation. Generally, the first-shell potential can be written as $U_m(r) = \sum_{j=1}^{3} A_j cos(\boldsymbol{G}_j \cdot r + \theta_j)$. The phases determine the real-space registry of the moiré potential. They are important for a microscopic miniband calculation, but the finite-$q$ electrostatic transfer from the top layer to the remote subsystem depends primarily on the magnitude of the Fourier component $|\delta n_G|$. Therefore, in the order-of-magnitude screening estimate below, we set $A_j = A$ and choose a common origin such that $\theta_j = 0$, following the simplified scalar model used by Yankowitz et al [7]. Here, $A$ is the amplitude of each first-shell cosine harmonic, and $\mathbf{G}_j$ $(j = 1,2,3)$ are the three non-redundant first-shell moiré reciprocal lattice vectors. We evaluate a range $A$ = 20–60 meV [7, 12] rather than relying on a single value.

We write the local Dirac-point energy of the hBN-aligned top graphene as $E_D(r) = \bar{E}_D + U_m(r)$, where $E_D(r)$ is the local Dirac-point energy at the in-plane position *r*, $\bar{E}_D$ is the spatially averaged Dirac-point energy in the absence of the moiré modulation, and $U_m(r)$ is the moiré potential generated by the hBN/graphene moiré superstructure.

The local Fermi energy measured from the local Dirac point is $\xi(r) = \tilde{\mu}_{MLG} - A\sum_{j=1}^{3} cos(\boldsymbol{G}_j \cdot r)$, where $\tilde{\mu}_{MLG}$ is the chemical potential measured from the spatially averaged Dirac point of the top graphene. Using the Dirac dispersion of monolayer graphene, the local carrier density is

$$n_{top}(r) = sgn[\xi(r)]\frac{\xi(r)^2}{\pi(\hbar v_F)^2}$$

$$= sgn\left[\tilde{\mu}_{MLG} - A\sum_{j=1}^{3} cos(\boldsymbol{G}_j \cdot r)\right]\frac{\left[\tilde{\mu}_{MLG} - A\sum_{j=1}^{3} cos(\boldsymbol{G}_j \cdot r)\right]^2}{\pi(\hbar v_F)^2}.$$

We denote this moiré-modulated chemical potential as $\tilde{\mu}_{MLG}$, to distinguish it from the uniform-density Fermi energy $\mu_{MLG} = \hbar v_F\sqrt{\pi|n_{MLG}|}$ used in the main manuscript. Since this density is periodic with the moiré wavelength, it can be expanded as $n_{top}(r) = \langle n_{top}\rangle + \sum_{G\in g_1} n_G e^{iG\cdot r} + \sum_{Q\in g_{>1}} n_Q e^{iQ\cdot r}$. Here $\langle n_{top}\rangle$ is the spatially averaged top-graphene density, corresponding to $n_{\mathrm{MLG}}$ in our experiment, $g_1 = \{\pm G_1, \pm G_2, \pm G_3\}$ is the first moiré reciprocal shell and $g_{>1}$ contains the higher-harmonics components such as $2G_j$ *and* $G_i - G_j$, which are generated by the nonlinear dependence of the carrier density on the local moiré potential. Because $n_{top}(r)$ is real, $n_{-G} = n_G^*$. Therefore, the first-shell harmonic terms can be written as $\sum_{G\in g_1} n_G e^{iG\cdot r}=\sum_{j=1}^{3} \delta n_j cos(\boldsymbol{G}_j \cdot r + \varphi_j)$, where $\delta n_j = 2\left|n_{G_j}\right|$ and $\varphi_j$ is the phase of the corresponding Fourier coefficient. Here, $n_{G_j}$ is the complex Fourier coefficient of the $+G_j$ component, while $\delta n_j$ is the real-space cosine amplitude obtained after combining the $+G_j$ and -$G_j$ components. Among these, the first-shell component is the most relevant source of the remote Hartree potential, because higher-wavevector components are more strongly suppressed during the finite-q electrostatic transfer.

In the simplified $C_3$-symmetric estimate, the density-modulation amplitudes associated with the three first-shell components are taken to be equal, $\delta n_1 = \delta n_2 = \delta n_3 \equiv \delta n_G$. Thus, $\delta n_G = 2|n_{G_1}| = 2|n_{G_2}| = 2|n_{G_3}|$.

Note that the special triangular-lattice relation $G_1 + G_2 + G_3 = 0$ is useful for evaluating the density modulation. When we introduce two dimensionless moiré coordinates as $x = G_1 \cdot r$, $y = G_2 \cdot r$, first-shell moiré potential becomes $U_m(r) = \sum_{j=1}^{3} A_j cos(\boldsymbol{G}_j \cdot r) = A[\cos x + \cos y + \cos(x+y)]$. This allows the unit-cell average to be performed by integrating x and y from 0 to 2π. With this notation, the local Fermi energy is $\xi(x,y) = \tilde{\mu}_{MLG} - A[\cos x + \cos y + \cos(x+y)]$, and the local carrier density becomes $n_{top}(x,y) = sgn[\xi(x,y)]\frac{\xi(x,y)^2}{\pi(\hbar v_F)^2}$. Using various value of $A$ = 20, 40, and 60 meV and $v_F = 1.12 \times 10^6$ m/s, we numerically determine $\tilde{\mu}_{MLG}$ from the condition that the spatial average of the local carrier density over one moiré unit cell equals $n_{MLG}$:

$$n_{MLG} = \frac{1}{(2\pi)^2}\int_0^{2\pi} dx \int_0^{2\pi} dy\, n_{top}(x,y) \; = \frac{1}{(2\pi)^2}\int_0^{2\pi} dx \int_0^{2\pi} dy\, sgn[\xi(x,y)]\, \frac{\xi(x,y)^2}{\pi(\hbar v_F)^2}.$$

With the determined $\tilde{\mu}_{MLG}$, we can obtain $n_{G_1} =< n_{top}(x,y)\, e^{-ix} >$ and $\delta n_G = 2\left|n_{G_1}\right|$. Table S3 and Fig. S10 summarize the chemical potential and density modulation at various $n_{MLG}$.

| $A$ (meV) | $n_{MLG}$ $(cm^{-2})$ | $\tilde{\mu}_{MLG}$ (meV) | $\delta n_G$ $(cm^{-2})$ |
|---|---|---|---|
| 20 | 0 | 4.10 | $4.00 \times 10^{10}$ |
| | $0.5 \times 10^{12}$ | 89.09 | $1.85 \times 10^{11}$ |
| | $1.0 \times 10^{12}$ | 128.35 | $2.77 \times 10^{11}$ |
| | $1.5 \times 10^{12}$ | 158.15 | $3.47 \times 10^{11}$ |
| 40 | 0 | 8.21 | $1.60 \times 10^{11}$ |
| | $0.5 \times 10^{12}$ | 79.04 | $2.88 \times 10^{11}$ |
| | $1.0 \times 10^{12}$ | 121.13 | $4.74 \times 10^{11}$ |
| | $1.5 \times 10^{12}$ | 152.35 | $6.20 \times 10^{11}$ |
| 60 | 0 | 12.31 | $3.60 \times 10^{11}$ |
| | $0.5 \times 10^{12}$ | 68.44 | $4.22 \times 10^{11}$ |
| | $1.0 \times 10^{12}$ | 109.76 | $6.00 \times 10^{11}$ |
| | $1.5 \times 10^{12}$ | 142.36 | $7.96 \times 10^{11}$ |

**Table S3| The chemical potential and density modulation as a function of $n_{MLG}$.**

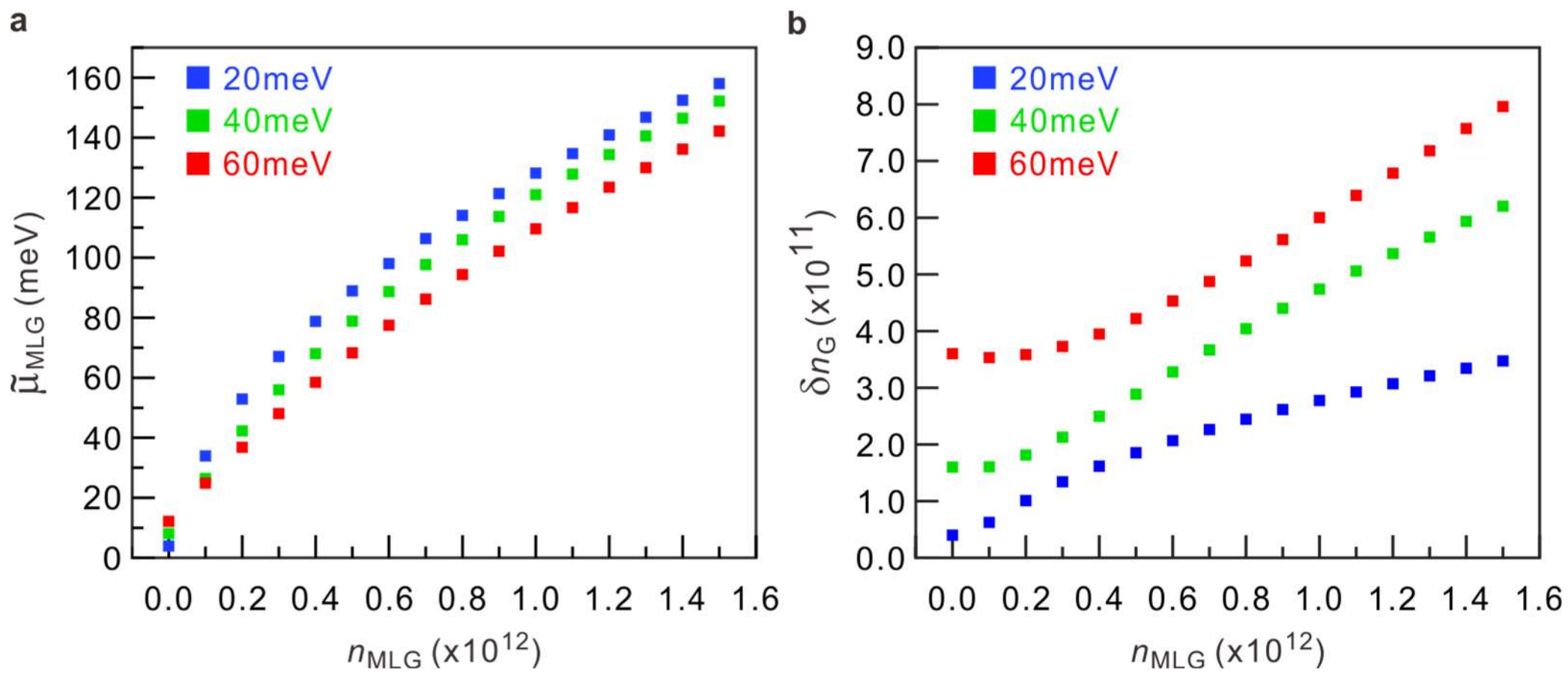

**Figure S10| (a) The chemical potential and (b) density modulation as a function of $n_{MLG}$ at various $A$.**

## Unscreened periodic Hartree potential and finite-q interlayer capacitance

We next estimate the electrostatic potential produced by the moiré-periodic density modulation in the top graphene. We take the charge density to be $\rho(r,z) = -e\sum_{j=1}^{3}\delta n_G cos(\mathbf{G}_j\cdot r)\delta(z)$, where $\rho(r,z)$ is the three-dimensional charge density and $z$ is the out-of-plane coordinate normal to the graphene plane. Here, we define $z = 0$ at the top graphene layer where the moiré-periodic charge-density modulation is located. The middle and bottom graphene layers are located at $z = d$ and $z = 2d$, respectively, so the center of the remote TBG subsystem is at $z = 1.5d$. This coordinate convention and electrostatic-transfer geometry are illustrated schematically in Fig. S11.

Solving the Poisson equation in the dielectric region between the graphene sheets gives $\phi(r,z) = \frac{-e\delta n_G}{2\epsilon_0\epsilon_{eff}G_m}\sum_{j=1}^{3}e^{-G_m z}cos(\mathbf{G}_j\cdot r)$, where $G_m = |\boldsymbol{G}_j|$ is common to all first-shell moiré reciprocal vectors. This expression shows that a finite-wavevector electrostatic potential decays exponentially along the out-of-plane direction with decay length $l_z = G_m^{-1}$. This is the relevant vertical finite-$q$ decay length.

For each first-shell component, the bare transferred potential energy at the TBG subsystem is $U_{TBG}(z) = \frac{e^2\delta n_G}{2\epsilon_0\epsilon_{eff}G_m}e^{-G_m z}cos(\mathbf{G}_j\cdot r) = U_{TBG}^{(0)}(z)cos(\mathbf{G}_j\cdot r)$, where $U_{TBG}^{(0)}(z)$ denotes the amplitude of one first-shell cosine component of the transferred periodic potential before including the self-screening of either the top MLG or the TBG subsystem.

For device M1, the moiré wavelength is $\lambda_m$ = 13.5 nm. For a triangular moiré lattice, $G_m = \frac{4\pi}{\sqrt{3}\lambda_m} = 0.53\ nm^{-1}$, so the vertical finite-q electrostatic decay length is $l_z = G_m^{-1} = 1.9\ nm$. This length is larger than the trilayer-graphene thickness. Using the density modulation $\delta n_G$ obtained in Table S3, taking the distance from the top graphene to the center of the TBG subsystem as z = 1.5d = 0.51 nm, and using $\varepsilon_{eff}$ = 3.5 extracted from the uniform-field interlayer capacitance $C_{int}$, we estimate $U_{TBG}^{(0)}(z)$; the results at various $n_{MLG}$ are summarized in Table S4.

| $A$ (meV) | $n_{MLG}\ (cm^{-2})$ | $U_{TBG}^{(0)}(z = 1.5d)$ (meV) |
|---|---|---|
| 20 | 0 | 1.5 |
| | $0.5 \times 10^{12}$ | 6.9 |
| | $1.0 \times 10^{12}$ | 10.3 |
| | $1.5 \times 10^{12}$ | 12.9 |
| 40 | 0 | 6.0 |
| | $0.5 \times 10^{12}$ | 10.7 |
| | $1.0 \times 10^{12}$ | 17.6 |
| | $1.5 \times 10^{12}$ | 23.1 |
| 60 | 0 | 13.4 |
| | $0.5 \times 10^{12}$ | 15.7 |
| | $1.0 \times 10^{12}$ | 22.3 |
| | $1.5 \times 10^{12}$ | 29.6 |

**Table S4| The bare transferred potential energy amplitude at $z$ = 1.5 $d$ without screening as a function of $n_{MLG}$.**

The same unscreened expression can be written in terms of a finite-q electrostatic capacitance:

$$U_{TBG}(z) = \frac{e^2 \delta n_G}{2\epsilon_0\epsilon_{eff}G_m} e^{-G_m z} cos(\mathbf{G}_j \cdot r) = \frac{e^2 \delta n_G}{C_G(z)} cos(\mathbf{G}_j \cdot r), C_G(z)$$
$$= 2\epsilon_0\epsilon_{eff}G_m e^{G_m z} .$$

For $z = 1.5d$, we obtain $C_G(z) \approx 4.4$ μF/cm$^2$. This is smaller than the uniform interlayer capacitance $C_{int} = \varepsilon_0\varepsilon_{eff}/z \approx 6.12$ μF/cm$^2$. The difference arises from the finite-$q$ electrostatic kernel exp($-G_m \cdot z$). Since the moiré wavelength is much larger than the interlayer spacing, the correction remains order-one: $C_G/C_{int} = 2G_m \cdot z \cdot \exp(G_m \cdot z) \approx 0.7$.

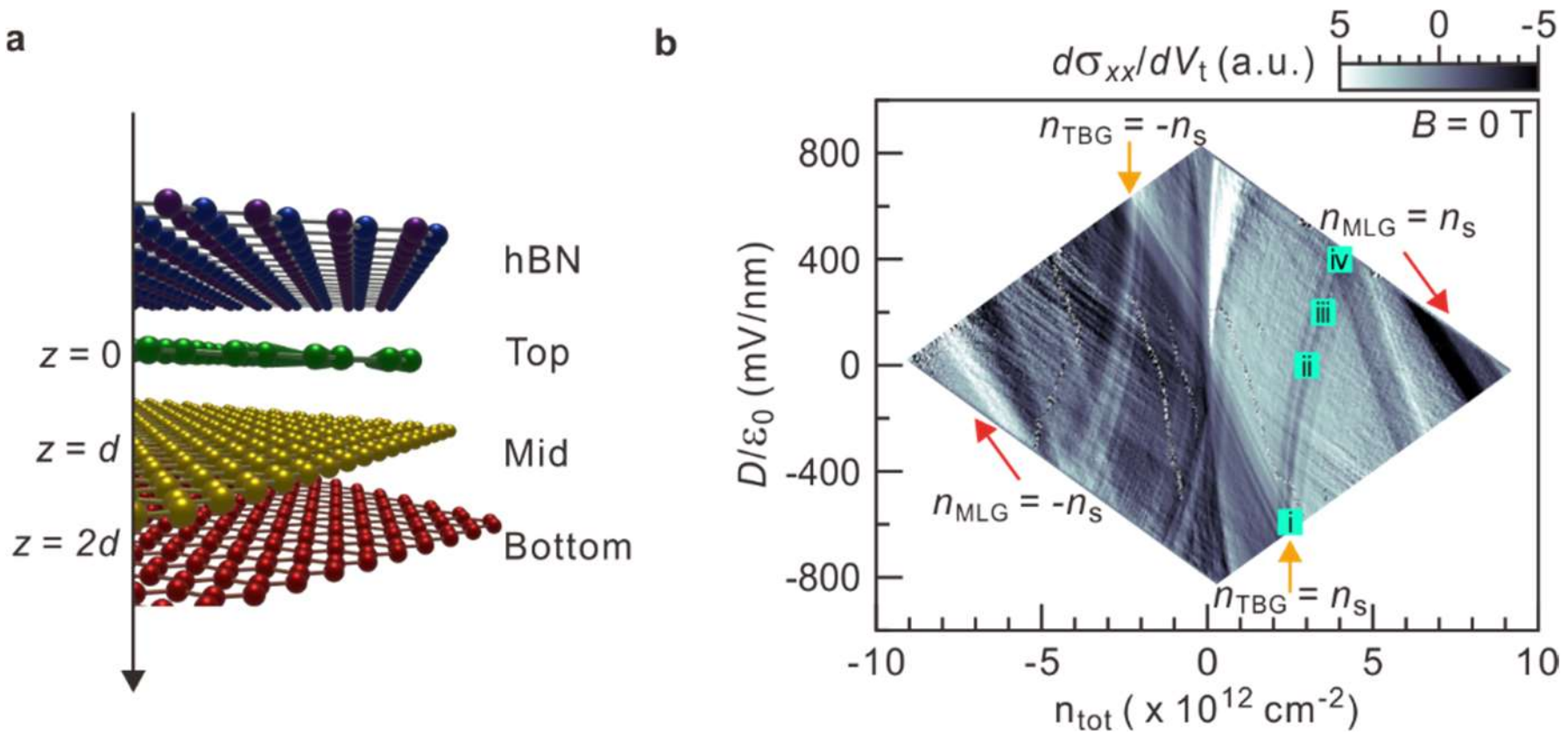

**Figure S11| Schematic of Moiré-modulated twisted trilayer graphene and Derivative of $\sigma_{xx}$ in ($n_{tot}$, $D/\varepsilon_0$) plane at $B$ = 0 T.** Red and Orange arrows indicate the secondary Dirac point of the hBN-aligned top graphene and the satellite features associated with the remote middle-bottom TBG subsystem. The cyan squares mark the four representative points at which we estimate the remote moiré potential amplitude: from ($n_{MLG}$, $n_{TBG}$) = (0, $n_s$) to (1.5 × $10^{12}$ cm$^{-2}$, $n_s$). Here, $n_s$ = 2.52 × $10^{12}$ cm$^{-2}$ is the moiré density corresponding to the secondary Dirac point.

**Periodic Hartree potential after TBG Thomas-Fermi screening**

The TBG subsystem does not experience the incoming periodic potential in its bare form. Before the potential reaches the remote TBG subsystem and is further screened by it, the finite-q Hartree potential generated by the top-layer density modulation is already screened by the electronic compressibility of the hBN-aligned top monolayer graphene. We therefore first estimate the top MLG screening from the spatially averaged local compressibility of the moiré modulated top graphene, rather than from the uniform-density graphene expression alone.

$$C_Q^{MLG}(n_{MLG}) = e^2 \frac{\delta n_{MLG}}{\delta \tilde{\mu}_{MLG}} = e^2 \langle \frac{\delta n_{top}(x,\ y)}{\delta \tilde{\mu}_{MLG}} \rangle = e^2 \langle \frac{2\xi(x,y)}{\pi(\hbar v_F)^2} \rangle$$

The corresponding Thomas-Fermi wavevector and finite-$q$ screening factor are $q_{TF}^{MLG}(n_{MLG}) = \frac{C_Q^{MLG}(n_{MLG})}{2\epsilon_0 \epsilon_{eff}}$ , $S_{MLG}(G_m) = \frac{G_m}{G_m + q_{TF}^{MLG}}$. The screening factors as a function of $n_{\mathrm{MLG}}$ are summarized in Table S5.

| $A$ (meV) | $n_{MLG}$ $(cm^{-2})$ | $C_Q^{MLG}(n_{MLG})$ | $q_{TF}^{MLG}(n_{MLG})$ | $S_{MLG}(G_m)$ |
|---|---|---|---|---|
| 20 | 0 | 0.41 | 0.066 | 0.890 |
| | $0.5 \times 10^{12}$ | 1.67 | 0.270 | 0.663 |
| | $1.0 \times 10^{12}$ | 2.41 | 0.389 | 0.577 |
| | $1.5 \times 10^{12}$ | 2.97 | 0.479 | 0.525 |
| 40 | 0 | 0.81 | 0.131 | 0.802 |
| | $0.5 \times 10^{12}$ | 1.56 | 0.252 | 0.678 |
| | $1.0 \times 10^{12}$ | 2.27 | 0.367 | 0.591 |
| | $1.5 \times 10^{12}$ | 2.86 | 0.461 | 0.535 |
| 60 | 0 | 1.22 | 0.197 | 0.729 |
| | $0.5 \times 10^{12}$ | 1.69 | 0.272 | 0.661 |
| | $1.0 \times 10^{12}$ | 2.21 | 0.357 | 0.598 |
| | $1.5 \times 10^{12}$ | 2.71 | 0.438 | 0.548 |

**Table S5| The quantum capacitance, Thomas-Fermi wavevector, and finite-*q* screening factor.**

This calculation shows that the top layer screening remains finite even at $n_{\mathrm{MLG}}$=0 when the moiré-modulated chemical potential $\tilde{\mu}_{MLG}$ and the spatially varying local Fermi energy ξ(x,y) are used. This finite screening arises because the moiré potential creates local electron and hole regions within the unit cell, making the local-compressibility average more appropriate than the uniform-density approximation, for which $C_Q^{MLG} = 0$ at charge neutrality.

The TBG subsystem itself also screens the transferred potential: the incoming perturbation induces a density modulation in the TBG, and this induced charge partially screens the original potential. Approximating the large-angle TBG as two monolayer-like Dirac cones gives

$$C_Q^{TBG} = e^2 \frac{\delta n_{TBG}}{\delta \mu_{TBG}} = \frac{2e^2}{\hbar v_f} \sqrt{\frac{2|n_{TBG}|}{\pi}}.$$

At the density where the satellite feature appears in device M1, $n_{\mathrm{TBG}} = n_{\mathrm{s}} = 2.52 \times 10^{12}$ cm$^{-2}$, we obtain $C_Q^{TBG} \approx 6.2$ μF/cm$^2$, $q_{\mathrm{TF}} \approx 0.99$ nm$^{-1}$, which gives the TBG finite-$q$ screening factor $S_{TBG}(G_m) = \frac{G_m}{G_m + q_{TF}^{TBG}} \approx 0.35$. The screened periodic potential acting on the TBG subsystem can be estimated as

$$U_{TBG}^{eff}(z) = S_{TBG}(G_m) S_{MLG}(G_m)\, U_{TBG}^{(0)}(z).$$

Using the $n_{\mathrm{MLG}}$-dependent $S_{MLG}(G_m)$ and the $S_{TBG}(G_m)$ evaluated at $n_{\mathrm{TBG}} = n_{\mathrm{s}}$, we estimate $U_{TBG}^{eff}(z)$ for the four representative points marked by squares in Fig. S11. The resulting screened potential amplitudes are summarized in Table S6.

| $A$ (meV) | Position | $n_{MLG}$ (cm$^{-2}$) | $S_{MLG}(G_m)$ | $n_{TBG}$ (cm$^{-2}$) | $S_{TBG}(G_m)$ | $U_{TBG}^{eff}(z)$ (meV) | $\frac{U_{TBG}^{eff}(z)}{A}$ (%) |
|---|---|---|---|---|---|---|---|
| 20 | i | 0 | 0.890 | $2.52 \times 10^{12}$ | 0.35 | 0.5 | 2.5 |
| | ii | $0.5 \times 10^{12}$ | 0.663 | $2.52 \times 10^{12}$ | 0.35 | 1.6 | 8.0 |
| | iii | $1.0 \times 10^{12}$ | 0.577 | $2.52 \times 10^{12}$ | 0.35 | 2.1 | 10.5 |
| | iv | $1.5 \times 10^{12}$ | 0.525 | $2.52 \times 10^{12}$ | 0.35 | 2.4 | 12.0 |
| 40 | i | 0 | 0.802 | $2.52 \times 10^{12}$ | 0.35 | 1.7 | 4.3 |
| | ii | $0.5 \times 10^{12}$ | 0.678 | $2.52 \times 10^{12}$ | 0.35 | 2.5 | 6.3 |
| | iii | $1.0 \times 10^{12}$ | 0.591 | $2.52 \times 10^{12}$ | 0.35 | 3.7 | 9.3 |
| | iv | $1.5 \times 10^{12}$ | 0.535 | $2.52 \times 10^{12}$ | 0.35 | 4.3 | 10.8 |
| 60 | i | 0 | 0.729 | $2.52 \times 10^{12}$ | 0.35 | 3.4 | 5.7 |
| | ii | $0.5 \times 10^{12}$ | 0.661 | $2.52 \times 10^{12}$ | 0.35 | 3.6 | 6.0 |
| | iii | $1.0 \times 10^{12}$ | 0.598 | $2.52 \times 10^{12}$ | 0.35 | 4.7 | 7.8 |
| | iv | $1.5 \times 10^{12}$ | 0.548 | $2.52 \times 10^{12}$ | 0.35 | 5.7 | 9.5 |

**Table S6| Screened first-shell periodic Hartree potential at $z$ = 1.5$d$ after including both top-MLG and TBG screening.**

Thus, after including both the top-MLG and TBG Thomas-Fermi screening, the transferred first-shell periodic potential is reduced to the few-meV scale. For $A$ = 20, 40, and 60meV, we estimate $U_{TBG}^{eff}(z = 1.5d)$ ≈ 0.5–2.4, 1.7-4.3, and 3.4-5.7 meV, respectively, over the representative points considered here. Equivalently, the effective remote potential corresponds to approximately 2.5-12% of the direct hBN/graphene moiré potential amplitude $A$, depending on $A$ and $n_{\mathrm{MLG}}$. The remote potential is therefore large enough to act as a weak periodic electrostatic perturbation on the middle–bottom subsystem, while remaining much weaker than the primary moiré potential in the top MLG. This few-meV scale is consistent with the satellite-like feature being much weaker than the primary hBN/graphene secondary-Dirac-point response. We do not attempt to fit the absolute transport intensity, which can also depend on disorder broadening, parallel conduction, contact geometry, and the density-dependent conductivity of each subsystem.

## Common-mode character of the remote moiré perturbation in the lower TBG subsystem

One possible alternative explanation is that the TBG-like lower-subsystem response arises simply from a small mismatch between the top–middle and middle–bottom twist angles.

However, previous angle-dependent calculations for relaxed twisted bilayer graphene further show that, above approximately 3°, the interlayer coupling changes only weakly with twist angle [13]; the observed subsystem organization is therefore unlikely to originate from a small mismatch between the top–middle and middle–bottom twist angles.

The combined lower-subsystem response is consistent with a weak remote moiré-periodic electrostatic perturbation sensed by the middle–bottom subsystem, and the uniform capacitance response and the finite-$G_m$ moiré response describe different aspects of the electrostatics. The potential generated by the top and bottom gates is a spatially uniform ($q$ = 0) field; in a capacitance-dominated response, independently resolved layers acquire different gate lever arms and therefore distinct slopes in the $n_{tot}$–$D/\varepsilon_0$ map, as observed in large-angle TTG devices without hBN alignment [4]. In contrast, the moiré-induced component considered here is a finite-$G_m$ periodic perturbation, whose amplitude at the two lower layers we now estimate.

The moiré potential generated by the top hBN/top graphene interface can be written as

$$U(r,z) = \frac{e^2 \delta n_G}{2\epsilon_0 \epsilon_{eff} G_m} e^{-G_m z} \sum_{j=1}^{3} cos(\boldsymbol{G}_j \cdot r + \theta_j),$$

where $\boldsymbol{G}_j$ are the three non-redundant first-shell moiré reciprocal-lattice vectors and $G_m = |\boldsymbol{G}_j|$. Comparing the potential felt by the bottom graphene layer at $z = 2d$ with that felt by the middle graphene layer at $z = d$, we obtain $U_{bottom}/U_{mid} = \exp(-G_m d) \approx 0.83$. Thus, the potential amplitudes on the middle and bottom graphene layers are not significantly different. The remote moiré potential therefore acts approximately as a long-wavelength periodic potential that is common to both lower graphene layers. The perturbation Hamiltonian due to $U_{mid}$ and $U_{bottom}$ can be written as

$$H_{pert} = \int d^2r \, [U_m(r) n_m(r) + U_b(r) n_b(r)].$$

We then introduce the symmetric and antisymmetric density modes, $n_+ = (n_m + n_b)$, $n_- = (n_m - n_b)$, and the corresponding conjugate potentials,

$$U_+(r) = \frac{U_m(r) + U_b(r)}{2}, \; U_-(r) = \frac{U_m(r) - U_b(r)}{2}.$$

The perturbation Hamiltonian can then be rewritten as

$$H_{pert} = \int d^2r\,[U_+(r)n_+(r) \;+\; U_-(r)n_-(r)].$$

Here, $U_+(r)$ couples to the total density mode $n_+$, whereas $U_-(r)$ couples to the density-difference mode $n_-$. In our geometry, $\frac{U_-}{U_+} = \frac{1 - e^{-G_m d}}{1 + e^{-G_m d}} \approx 0.09$. Therefore, more than 90% of the perturbation is in the symmetric component $U_+(r)$, while the antisymmetric component $U_-(r)$ is less than 10%. Since $G_m d < 1$, the finite-$G_m$ electrostatic perturbation changes only weakly between the middle and bottom graphene layers. Thus, the remote moiré field is expected to reach both lower layers with comparable amplitude and to generate only a small differential potential between them. However, a nearly common scalar potential does not by itself hybridize the two layers or prove a collective TBG miniband. Our transport measurement resolves the combined middle-bottom response, but it does not determine the layer-by-layer distribution of the induced modulation.

The MLG-like top-layer response, the lower-subsystem satellite feature locked to $n_{TBG} \approx \pm n_s$, and the finite-$G_m$ electrostatic estimate consistently point to a weak remote moiré-periodic electrostatic modulation sensed by the middle–bottom subsystem.

## Supplementary References